\documentclass{article}
\usepackage{graphicx}
\usepackage{epstopdf, epsfig}
\usepackage{graphics}
\usepackage{subfigure}
\usepackage{amsmath}
\usepackage{verbatim}
\usepackage[sort&compress,square,comma,authoryear]{natbib}
\title{Boundary layer in linear viscoelasticity}
\date{}

\author{Hualong Feng\thanks{hfeng@csub.edu}\\ \\Department of Mathematics, California State University, Bakersfield\\9001 Stockdale Highway, Bakersfield, CA 93311}

\newcommand{\Wi}{\textit{Wi}}
\newcommand{\Db}{\mathbf{D}}
\newcommand{\Fb}{\mathbf{F}}

\newcommand{\fb}{\mathbf{f}}
\newcommand{\nb}{\mathbf{n}}
\newcommand{\qb}{\mathbf{q}}

\newcommand{\vb}{\mathbf{v}}

\newcommand{\tb}{\mathbf{t}}
\newcommand{\xb}{\mathbf{x}}

\newcommand{\xbnot}{\mathbf{x}_0}
\newcommand{\sfi}{\int_{\partial\Omega}}

\begin{document}

\maketitle

\begin{abstract}
It is well known that a boundary layer develops along an infinite plate under oscillatory motion in a Newtonian fluid. In this work, this oscillatory boundary layer theory is generalized to the case of linear viscoelastic (LVE) flow. We demonstrate that the dynamics in LVE are generically different than those for flow of similar settings in Newtonian fluids, in several aspects. These new discoveries are expected to have consequences on related engineering applications. Mimicking the theory for Stokes oscillatory layers along an infinite plate in Newtonian flow, we derive a similar oscillatory boundary layer formula for the case of LVE. In fact, the new theory includes the Stokes layer theory as a special case. For the disturbance flow caused by particles undergoing oscillatory motion in linear viscoelasticity (LVE), a numerical investigation is necessary. A boundary integral method is developed for this purpose. We verify our numerical method by comparing its results to an existing analytic solution, in the simple case of a spherical particle. Then the numerical method is applied in case studies of more general geometries. Two geometries are considered because of their prevalence in applications: spheroids; dumbbells and biconcave disks.
\end{abstract}

{\bf \quad\ Key words:} Boundary layer, Oscillatory flow, LVE, boundary integral method, eddies.\\

\section{Introduction}
Oscillatory flows find applications in numerous fields of engineering, e.g., chemical engineering and biomedical engineering. The phenomena have been extensively studied theoretically, experimentally, and numerically \citep{Hall_1974, Sobey_1980, Stephanoff_1980, Sobey_1982, Sobey_1983, Sobey_1985_1, Sobey_1985_2, Ralph_1986, Ghaddar_1986a, Ghaddar_1986b, Pedley_1985, Padmanabhan_1987, Riley_1967, Pienkowska_1984}. 
Important non-dimensional groups in those studies include the Strouhal and the Reynolds number, hence the dynamics are investigated for various values of those numbers. Researchers also seek to under enhancement of mass and heat transport, often enabled by oscillatory flow. Other topics of interest include calculation of torque and stress on those particles, and analysis of hydrodynamic instabilities triggered by oscillatory flow.

It is known \citep{poz89} that in unsteady Stokes flow, a single frequency oscillatory motion of a sphere in a Newtonian fluid causes the appearance of an eddy around the particle. In each period, an eddy arises on the sphere and propagates away from the particle surface, and disappears in the interior of the fluid, in the plane perpendicular to the motion of the particle. This feature is absent when the inertial effect can be neglected (low frequency motion), because the flow is close to being Stokes in this regime. In many applications in both materials science and biology (or biophysics), the media are usually modeled as Newtonian, but they are typically often viscoelastic, i.e. they are both viscous and elastic depending on what time scales we observe them. Assuming them to be Newtonian is not because it is known to be a good assumption, but rather because accounting for viscoelasticity is difficult.

In this work, we study oscillatory motion of particles in viscoelastic fluids. When the particle undergoes small amplitude motion, linear viscoelasticity (LVE) applies, even if the frequency of the motion is high. In other words, LVE is valid when the strain is small, though the strain rate may be high. In fact, the high strain rate case is of interest here since if both the strain and strain rate are small, the motion in viscoelastic fluids reduces to that in Newtonian flow. We show that when viscoelastic effect can not be ignored (the Weissenberg number $\Wi$ is large), eddies are created around particles undergoing oscillatory motion, just like in the Newtonian flow case. However, there are significant differences from the Newtonian flow case.

For the purpose of numerical studies, we adopt a boundary integral method, due to its computational efficiency resulting from dimension deduction. Our method, however, is sufficient for the geometries considered in this work. For more complicated geometries, a boundary element method may be more appropriate. Our numerical method is globally third-order accurate. In a boundary integral or element method, a relation is established between the stress and velocity on particle surfaces, and then one quantity is solved given the other. This results in a Fredholm integral equation of the first or second kind. In our study, velocity of the particles is prescribed, and then the stress on the particle surfaces is solved for, which is in turn used to obtain the velocity in the interior of the fluid. This results in an ill-posed first-kind Fredholm integral equation. But we note that the right-hand side of the integral equation is prescribed and hence contains no error, thus the numerical errors due to the ill-conditioning will not compromise our solutions.

In Section~\ref{sec.me}, we present the mathematical equations used in this paper. We also write down the equations in the Fourier domain, since we are studying oscillatory flow. In Section~\ref{sec.bif}, we derive a boundary integral formulation for the mathematical equations contained in the first section, and for the particular applications we bear in mind. We also indicate how integrals can be performed along the azimuthal direction to convert the formulation to axisymmetric setting. In Section~\ref{sec.nm}, we illustrate our numerical method, which was explained with more details in a previous work. In Section~\ref{sec.r}, we present and discuss our results. We conclude the work in Section~\ref{sec.c}.

\section{Mathematical equations}\label{sec.me}

We will consider the oscillatory motion of solid particles suspended in an unsteady Stokes or linear viscoelastic (LVE) flow. First, we recall the linearized incompressible Cauchy equation without body force,
\begin{flalign}
\rho\dfrac{\partial\vb}{\partial t}=\nabla\cdot\boldsymbol{\tau}-\nabla p+\Fb,\quad\nabla\cdot\vb=0,
\label{eq.gen}
\end{flalign}
where $\rho$ is the fluid density, $\vb$ is the fluid velocity, $p$ is the pressure, $\Fb$ is the body force, and $\boldsymbol{\tau}$ is the extra stress tensor. Since we will only consider oscillatory motion caused by external force, the body force will be ignored. The no-slip boundary condition is imposed on the particle surfaces. For a Newtonian fluid, the stress tensor $\boldsymbol{\tau}=2\mu\Db$, where $\mu$ is the fluid dynamic viscosity and $\Db=(1/2)(\nabla\vb+\nabla\vb^T)$ is the symmetrized strain rate. Then, equation (\ref{eq.gen}) becomes
\begin{flalign}
\rho\dfrac{\partial\vb}{\partial t}=\mu\nabla^2\vb-\nabla p,\quad\nabla\cdot\vb=0.
\label{eq.stokes}
\end{flalign}
The stress tensor in a linear viscoelastic fluid is given by
\begin{flalign}
\boldsymbol{\tau}(\xb,t)=2\int_{-\infty}^tG(t-t')\Db(\xb,t')dt',
\label{eq.cons}
\end{flalign}
where $G(\cdot)$ is the relaxation modulus. For a Maxwell fluid with $N$ relaxation modes with a Newtonian component, $G$ reads
\begin{eqnarray}
G(t)=\mu\delta(t)+\sum\limits_{j=0}^{N-1}G_je^{-t/\lambda_j},
\label{eq:GN}
\end{eqnarray}
where $\mu$ is the dynamic Newtonian viscosity, $\delta$ is the Dirac's delta function, $\lambda_j$ and $G_j$ are the relaxation time and spectral strength of mode $j$, respectively. The magnitude of $\lambda_j$ controls how fast a certain mode relaxes, and the largest $\lambda_j$ is often used as the time scale when non-dimensionalizing the dynamics. With the formulation Eq.~(\ref{eq:GN}), any motion near equilibrium in a viscoelastic fluid can be approximated to an arbitrary accuracy using enough modes, {\it i.e.} by increasing $N$. For illustration, we take a one-mode Maxwell fluid with relaxation modulus $G(t)=G_0 e^{-t/\lambda_0}$, then the constitutive equation (\ref{eq.cons}) becomes
\begin{eqnarray}
\boldsymbol{\tau}(\xb,t)=\int_{-\infty}^t G_0 e^{(t-t')/\lambda_0}(\nabla\vb+\nabla\vb^T)dt'.
\label{eq.oneint}
\end{eqnarray}
There is a well known correspondence principle that relates the linear viscoelastic equation to the unsteady Stokes equation, in the Fourier domain. We take the time Fourier transform of Eq.~(\ref{eq.gen}) to obtain
\begin{flalign}
\rho i\omega\hat\vb= \hat G(\omega)\nabla^2\hat\vb-\nabla \hat p,\quad\nabla\cdot\hat\vb=0,
\label{eq.gfreq}
\end{flalign}
where $\hat\vb$ and $\hat p$ depend on space variable $\xb$ and frequency $\omega$. For a Newtonian fluid (unsteady Stokes flow), $\hat G(\omega)=\mu$, independent of $\omega$; while for a viscoelastic fluid (LVE), $\hat G (\omega)$ is the time Fourier transform of $G(t)$, and hence is $\omega$ dependent. Note that $G^*(\omega)=i\omega\hat G (\omega)$ is the dynamic modulus \citep{Schieber_13}, commonly used in characterizing dynamics under vibratory motion. We pick a characteristic length scale $l$, then Eq.~(\ref{eq.gfreq}) can be non-dimensionalized with a single parameter $\lambda^2=i\omega l^2\rho/\hat G(\omega)$,
\begin{flalign}
(\nabla^2-\lambda^2)\vb=\nabla p,\quad\nabla\cdot\vb=0,
\label{eq.Brink}
\end{flalign}
where all the hats have been dropped for simplicity. For a Newtonian fluid with dynamic viscosity $\mu$, the nondimensional number $\lambda^2=i\omega l^2\rho/\mu$ is called the frequency parameter, and it measures the effect of inertia. We note that $\lambda^2$ is purely imaginary in the case of a Newtonian fluid. For a one-mode Maxwell fluid,
\begin{flalign}
\hat G (\omega)=\dfrac{G_0\lambda_0}{1+i\omega\lambda_0},
\end{flalign}
so
\begin{flalign}
\lambda^2=\dfrac{i\omega l^2\rho(1+i\omega\lambda_0)}{G_0\lambda_0}.
\label{para}
\end{flalign} 
Now $\lambda^2$ becomes fully complex, and contains two nondimensional groups. The first is the magnitude $|\lambda^2|$, related to the inertial effect. For convenience, we will call both $\lambda^2$ and $|\lambda^2|$ the frequency parameter. We denote the frequency parameter by $\beta$, and whether it refers to $\lambda^2$ or $|\lambda^2|$ should be clear from the context. The second nondimensional group is the Weissenberg number $\Wi=\omega\lambda_0$, related to elastic effect.

The time domain equation (\ref{eq.stokes}) becomes the Brinkman equation (\ref{eq.Brink}), in the frequency domain. Being a linear equation, it admits numerous well studied numerical methods. In particular, we will outline a boundary integral formulation in the next section to numerically solve Eq.~(\ref{eq.Brink}). We note that solving the equation in the frequency domain is sufficient, since we will be studying applications of oscillatory flow. Once the velocity profile is solved for in the frequency domain, it will be converted back to the time domain, so that the stream function can be constructed to study the flow around the particles.

\section{Boundary layer theory for linear viscoelasticity}\label{sec.blt}
In this section, we derive a boundary layer theory for flow past an infinite plate in linear viscoelasticity. In Section~\ref{sec.r}, we will validate our numerical results against this theory.

We solve equation (\ref{eq.gen}) in the upper half plane, with $(x,z)$ coordinates and $z$ being the vertical one. Without loss of generality, we assume there is no body force. An infinite plane at $z=0$ is undergoing horizontal oscillatory motion, furnishing one boundary condition of velocity. We assume there is no motion at infinity, so $\vb(x,\infty,t)=0$ is the other boundary condition of velocity. If we write out the $z$ component in equation (\ref{eq.gen}) and analyze the dynamics, we conclude $\partial p/\partial z=0$ since there is no motion in the $z$ direction (note that unlike the case of nonlinear viscoelasticity, there is no normal stress in LVE). As we will study oscillatory flow, we work in the frequency domain and solve equation (\ref{eq.Brink}) in the $z$ direction (so the $\nabla p$ term vanishes):
\begin{flalign}
(\nabla^2-\lambda^2)v_x=0,
\end{flalign}
subject to the boundary condition $v_x(0)=1$ and $v_x(\infty)=0$. The solution for the one-Maxwell fluid with the frequency parameter as shown in equation (\ref{para}) is
\begin{flalign}
v_x(z)=\exp\left(\sqrt{\dfrac{\rho\omega}{2G_0\lambda_0}}\left(-\sqrt{\sqrt{\Wi^2+1}+\Wi}+i\sqrt{\sqrt{\Wi^2+1}-\Wi}\right)z\right),
\label{sol}
\end{flalign}
where $\Wi=\omega\lambda_0$ is the Weissenberg number. For $\Wi\ll1$, the solution approaches the well known result in Stokes layer theory for Newtonian fluids, $v_x(z)=\exp\left(\sqrt{\omega/2\nu}(-1+i)z\right)$, where $\nu$ is the kinematic viscosity \citep[Chap. 5]{Batchelor_FD}. To see this, note that $G_0\lambda_0/\rho\to\mu/\rho:=\nu$, the kinematic viscosity for a Newtonian fluid, as the viscoelastic relaxation time $\lambda_0\to0$. We are interested in the case $\Wi\sim O(1)$ or $\Wi\gg1$. When $\Wi\gg1$, the solution (\ref{sol}) approaches $v_x(z)=\exp\left(-\omega\sqrt{\rho/G_0}z\right)$.

\begin{flalign}
v_x(z)=\exp\left(\sqrt{\dfrac{\rho\omega(G_0\lambda_0+\mu+\mu\omega^2\lambda_0^2)}{2[(\mu\omega\lambda_0)^2+(G_0\lambda_0+\mu)^2]}}\left(-\sqrt{\sqrt{w_p^2+1}+w_p}+i\sqrt{\sqrt{w_p^2+1}-w_p}\right)z\right),
\end{flalign}
where
$$w_p=\dfrac{G_0\omega\lambda_0^2}{G_0\lambda_0+\mu+\mu\omega^2\lambda_0^2}.$$

There are two differences between this asymptotic behavior and the case for Newtonian fluids. First, the boundary layer thickness scales like $O(\omega^{-1})$ at high values of $\omega$, while the thickness scales like $O(\omega^{-1/2})$ for the Stokes layer. Second, there is a phase difference of $\pi/2$ between the stress and the velocity on the horizontal plate, while the phase difference is $\pi/4$ for the Stokes layer \cite{Batchelor_FD}.

\section{Boundary integral formulation}\label{sec.bif}
\begin{figure}
\centering
\includegraphics[height=1.5in]{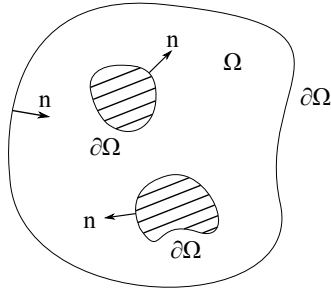}
\caption[]{Schematic for boundary integral formulation.}
\label{fig_bif}
\end{figure}
In this section, we will explain a boundary integral method to solve Eq.~(\ref{eq.Brink}). Following \citet{poz89,poz92}, the velocity $\vb e^{i\omega t}$ at a point $\xbnot$ on the surface $\partial\Omega$ satisfies the following boundary integral equation (BIE) 
\begin{eqnarray}
v_j(\xbnot)
  = - \frac{1}{4\pi}\sfi f_i(\xb) G_{ij}(\xb,\xbnot)\,{\rm d}S(\xb)
    + \frac{1}{4\pi}\sfi v_i(\xb) T_{ijk}(\xb,\xbnot) n_k(\xb)\,{\rm d}S(\xb),
\label{eq.bie1}
\end{eqnarray}
where $\fb e^{i\omega t}$ is the traction on the surface, the normal vector $\nb$ points into the fluid, and the integrations are principal value integrals. See Fig. \ref{fig_bif} for an illustration. We refer to $\xb$ and $\xbnot$ as the source and target point, respectively.The expressions of the Green's functions $G_{ij}$ and $T_{ijk}$ for Brinkman equation~(\ref{eq.Brink}) are given in \citet{poz89,poz92}.

The velocity of and the traction on the particles are related to each other according to the BIE (\ref{eq.bie1}), so one can solve for the traction if the velocity is given and vice versa. The advantage of this boundary integral method is that once the velocity and traction are known for the particles, the flow quantities in the whole fluid interior can be evaluated with simple integrals. Therefore, there is dimension reduction and hence saving in computational cost with this method. Eq~(\ref{eq.bie1}) is valid when the target point $\xbnot$ is on the surface. When $\xbnot$ is in the interior of the fluid, the boundary integral equation should read
\begin{eqnarray}
v_j(\xbnot)
  = - \frac{1}{8\pi} \sfi f_i(\xb) G_{ij}(\xb,\xbnot)\,{\rm d}S(\xb)
    + \frac{1}{8\pi} \sfi v_i(\xb) T_{ijk}(\xb,\xbnot) n_k(\xb)\,{\rm d}S(\xb).
\label{eq.biei}
\end{eqnarray}

Under certain circumstances, one can bypass the double layer integral in Eq~(\ref{eq.bie1}) to save computational cost. To this end, we introduce a ghost flow $\vb'$ inside the solid particles. W require that $\vb'$ satisfy the same boundary condition as $\vb$ on the surface and tend to 0 at infinity. Then, for a point $\xbnot$ in the interior of the fluid, we have \citep{poz89,poz92}
\begin{eqnarray}
 v_j(\xbnot) 
  = - \frac{1}{8\pi} \sfi q_i(\xb) G_{ij}(\xb,\xbnot)\,{\rm d}S(\xb), 
\label{eq.bie2} 
\end{eqnarray}
where $\qb := \fb - \fb'$ is an unknown surface density. The success of this simplification all hinges on if one can easily find the pair $(\vb',\fb')$. This is possible, at least wen the particles are not deformable. For example, when a single spherical particle undergoes a single frequency oscillation $\vb e^{i\omega t}$ ($\vb$ is a constant), we have \citep{kk1991}
\begin{eqnarray}
\vb'=\vb,\quad\fb'=\lambda^2 (\xb \cdot \vb) \nb,
\label{eq.exact1}
\end{eqnarray}
where $\nb$ is the unit outward normal to the surface. We note that equation \eqref{eq.bie2} is valid when $\xbnot$ is inside the fluid or on the particle surface. 


\begin{figure}
\centering
\includegraphics[height=2in]{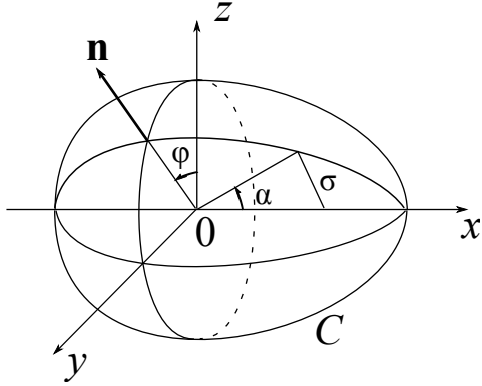}
\caption[]{Schematic of an axisymmetric boundary.}
\label{fig_axis}
\end{figure}
Since we will only study axisymmetric motion (with no swirl) of particles, we use cylindrical coordinates $\xb=(x,\sigma,\psi)$ and integrate along the azimuthal ($\psi$) direction and reduce the surface integral in Eq.~(\ref{eq.bie2}) to a line integral along the contour $C$ (semi-circle) of each particle in the $x-\sigma$,
\begin{eqnarray}
 v_i(\xbnot) = - \frac{1}{8\pi\mu} 
        \int_C M_{ij}(\xb,\xbnot) q_j(\xb) \,{\rm d}s(\xb), 
\label{eq.bie3}
\end{eqnarray}
where $i,j = 1,2$ indicate the axial $x$ and radial ($\sigma$) directions respectively, and $\textbf{M}$ is the modified Green's function tensor under cylindrical coordinates. The axisymmetric boundary integral formulation \eqref{eq.bie3} also works for the case of two or multiple particles if the definition of the boundaries $C$ includes all interfaces. Let $\xb=(x,\sigma,\psi)$ and $\xbnot=(x_0,\sigma_0,\psi_0)$ in cylindrical coordinates, then one calculates
\begin{eqnarray}
r=|\xb-\xbnot|=\sqrt{(x-x_0)^2+\sigma^2+\sigma_0^2-2\sigma\sigma_0\cos\phi},\quad\mbox{where }\phi=\psi-\psi_0.
\label{eq.dist}
\end{eqnarray}
Letting $\hat{x}=x-x_0$, the axisymmetric Green's function tensor $\textbf{M}$ in Eq.~(\ref{eq.bie3}) will be a function of $\hat{x}$ and $r$. Following \citet{poz89} and letting $\hat{x}=x-x_0$, the axisymmetric Green's function tensor $\textbf{M}$ in Eq.~(\ref{eq.bie3}) is
\begin{eqnarray}
 \textbf{M} = \sigma \left( \begin{array} {cc}
     A_{10} + \hat{x}^2 B_{30} & \hat{x} (\sigma B_{30} - \sigma_0 B_{31}) \\
     \hat{x} (\sigma B_{31} - \sigma_0 B_{30}) &
     A_{11} + (\sigma^2+\sigma_0^2) B_{31} - \sigma \sigma_0 (B_{30} + B_{32})
     \end{array} \right),
\label{eq.gf3}
\end{eqnarray}
where 
\begin{eqnarray}
 A_{mn} = \int_0^{2\pi} \frac{\cos^n \phi}{r^m} A(\lambda r) \, {\rm d}\phi,\quad
 B_{mn} = \int_0^{2\pi} \frac{\cos^n \phi}{r^m} B(\lambda r) \, {\rm d}\phi.
\label{eq.AB}
\end{eqnarray}

There is an eigen-solution, the normal vector $n_i({\bf x})$ of the particle surface, to the integral equation~(\ref{eq.bie2}) because of the identity
\begin{flalign}
\sfi n_i(\xb) G_{ij}(\xb,\xbnot)\,{\rm d}S(\xb)=0.
\label{eq.bie4}
\end{flalign}
On the other hand, physical considerations dictate that the solution $\qb$ to Eq.~(\ref{eq.bie2}) be orthogonal to $\nb$,
\begin{flalign}
\sfi \qb\cdot\nb \,{\rm d}S(\xb)=0.\label{eq.cond}
\end{flalign}
A solution to equation~(\ref{eq.bie2}) that satisfies the physical condition Eq.~(\ref{eq.cond}) can be obtained by solving instead the following equation \citep{LP01}
\begin{flalign}
 v_j(\xbnot) 
  =-\frac{1}{8\pi\mu} \sfi q_i(\xb) \left[G_{ij}(\xb,\xbnot) +
      n_j(\xbnot) n_i(\xb)\right]\,{\rm d}S(\xb).
\label{eq.re}
\end{flalign}
For an axisymmetric flow, it reads
\begin{flalign}
 v_i(\xbnot) = - \frac{1}{8\pi\mu} 
        \int_C \left[M_{ij}(\xb,\xbnot) 
     + n_i(\xbnot) n_j(\xb)\right] q_j(\xb) \,{\rm d}S(\xb),
\label{eq.reaxis} 
\end{flalign}
where $i,j$ now indicate axial and radial directions. Thus, if the velocity $\vb$ is given,  Eq.~(\ref{eq.reaxis}) provides a Fredholm integral equation of the first kind for the unknown surface density $\qb$.

\section{Numerical methods}\label{sec.nm}
A boundary element method to solve Eq.~(\ref{eq.reaxis}) is explained in \citet{poz89}, and the method is second order accurate in space. But we will use a boundary integral method \citep{LP01} that is third order accurate in space. Since Eq.~(\ref{eq.reaxis}) is ill-posed, the method \citep{LP01} is carefully designed to ensure that reliable solutions be obtained. 

We summarize how Eq.~(\ref{eq.reaxis}) is solved, with details contained in \citet{LP01}. First, the integrals $\{A_{mn},B_{mn}\}$ (being periodic in the azimuthal angel $\phi$) in Eq.~(\ref{eq.AB}) are not amenable to a straightforward application of the trapezoidal rule, and a hybrid quadrature was designed to improve the efficiency. Second, the integral equation (\ref{eq.reaxis}) contains a singular integral. Each entry of the Green's function tensor $\textbf{M}$ in Eq.~(\ref{eq.gf3}) is a function of $(\alpha, \alpha_l)$, the polar angles corresponding to the two points $\xb$ and $\xbnot$ in the meridional plane. $\textbf{M}(\alpha, \alpha_l)$ has a logarithmic singularity at $\alpha=\alpha_l$, so each entry $H$ in $\textbf{M}$ has the following expansion,
\begin{eqnarray}
H(\alpha,\alpha_l) = \tilde H(\alpha,\alpha_l) + \sum c_k(\alpha_l)(\alpha-\alpha_l)^k\log |\alpha-\alpha_l|,
\label{eq.asym}
\end{eqnarray}
where $\tilde H$ is smooth in $\alpha$. To account for the log singularity, the modified trapezoidal rule \citep{Sidi-1,Nitsche} is used because of its simplicity. It can be shown that
\begin{eqnarray}
\int_a^bH(\alpha,\alpha_l)d\alpha&=&h\sum\limits_{\substack{k=0\\k\neq l}}^N{'}H(\alpha_k,\alpha_l)+h\tilde H(\alpha_l,\alpha_l)+c_0(\alpha_l)h\log\dfrac{h}{2\pi}\\
&+&\sum\limits_{\substack{k=1\\k\scriptsize\mbox{ odd}}}^m \gamma_k(H^{(k)}(b,\alpha_l)-H^{(k)}(a,\alpha_l))h^{k+1}\nonumber+\sum\limits_{\substack{k=2\\k\scriptsize\mbox{ even}}}^m \nu_kc_k(\alpha_l)h^{k+1}+O(h^{m+2})
\label{eq.quad}
\end{eqnarray}
for $m\geq0$, where $h=(b-a)/N$, $\alpha_k=a+kh$, and the prime on the summation indicates that the two end summands are weighted by half. We use a version of the quadrature (\ref{eq.quad}) with $m=1$, so it will be third order accurate. The only constant in (\ref{eq.quad}) relevant here is $\gamma_1=-1/12$. The smooth part $\tilde H$ is interpolated to second order accuracy in $h$. Third, when the stress on particles is solved for, the accuracy order in the numerical result is not uniform along the meridional plane. Specifically, the error is third order away from the two poles ($\alpha=0$) on the axisymmetric surface, but only first order near the poles. A pole correction scheme was implemented to overcome this difficulty, resulting in a uniform third order accurate method.

A motion of the particle is prescribed, then GMRES \citep{Saad-1} is used to solve Eq.~(\ref{eq.reaxis}) for $\qb$ on the particle surface. Eq.~(\ref{eq.reaxis}) is again used to obtain velocities in the interior of the fluid. Since the flow is incompressible, a stream function $\psi$ is defined such that
\begin{flalign}
v_\sigma=-\dfrac{1}{\sigma}\dfrac{\partial\psi}{\partial x},\quad v_x=\dfrac{1}{\sigma}\dfrac{\partial\psi}{\partial\sigma}.
\label{eq.stream1} 
\end{flalign}
The stream function $\psi$ can be evaluated with the following path integral,
\begin{flalign}
\psi=\displaystyle{\int(\sigma v_xd\sigma-\sigma v_\sigma dx)}.
\label{eq.stream2} 
\end{flalign}
The Simpson's rule is used to evaluate the integral for $\psi$. Since $\qb=\fb-\fb'$ and $\fb'$ is known to us from Eq.~(\ref{eq.exact1}), the traction $\fb$ on the particle can be obtained. The shear force $f=\fb\cdot\tb$, where $\tb$ is the tangent vector to the surface contour, will be analyzed in the result section, since it gives insight into the flow around the particles. Note that in the time domain, the shear force is the real part of $fe^{i\omega t}=ae^{i\theta}e^{i\omega t}$, where $a$ and $\theta$ are the magnitude and phase angle of the shear force -- in the Fourier domain -- respectively, so one convenient way to visualize the shear force is to plot $a$ and $\theta$ along the particle contour.

\section{Results}\label{sec.r}
In this section, we present numerical results for the disturbance flow caused by a particle undergoing oscillatory motion in linear viscoelasticity, and compare them to those for unsteady Stokes flow. We will study the shear stress on the particle surface, and streamlines around them at representative times.

\subsection{Spheroids}
\begin{figure}
\centering
\hskip -0.2in
\subfigure[]{\includegraphics[height=1.3in]{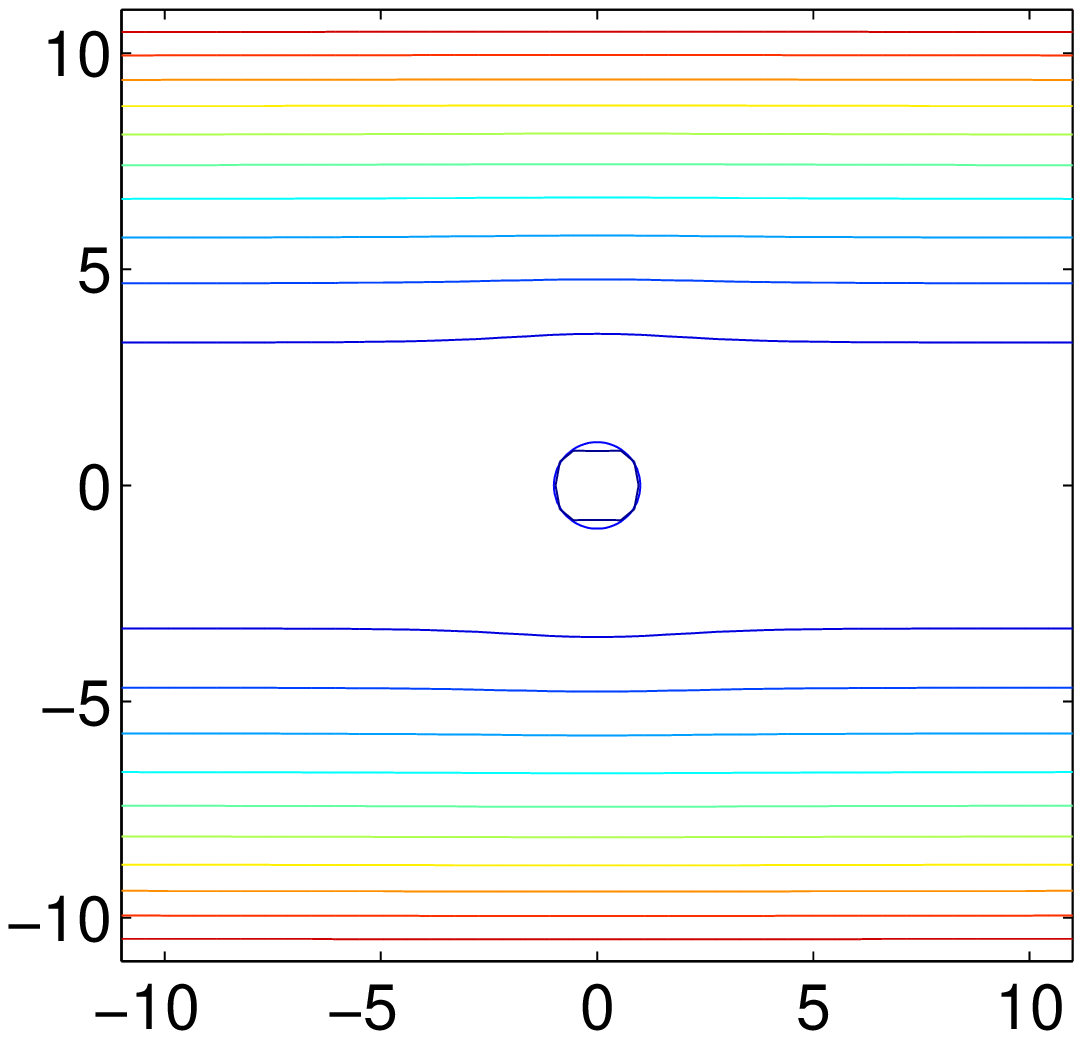}}
\hskip -0.3in
\subfigure[]{\includegraphics[height=1.3in]{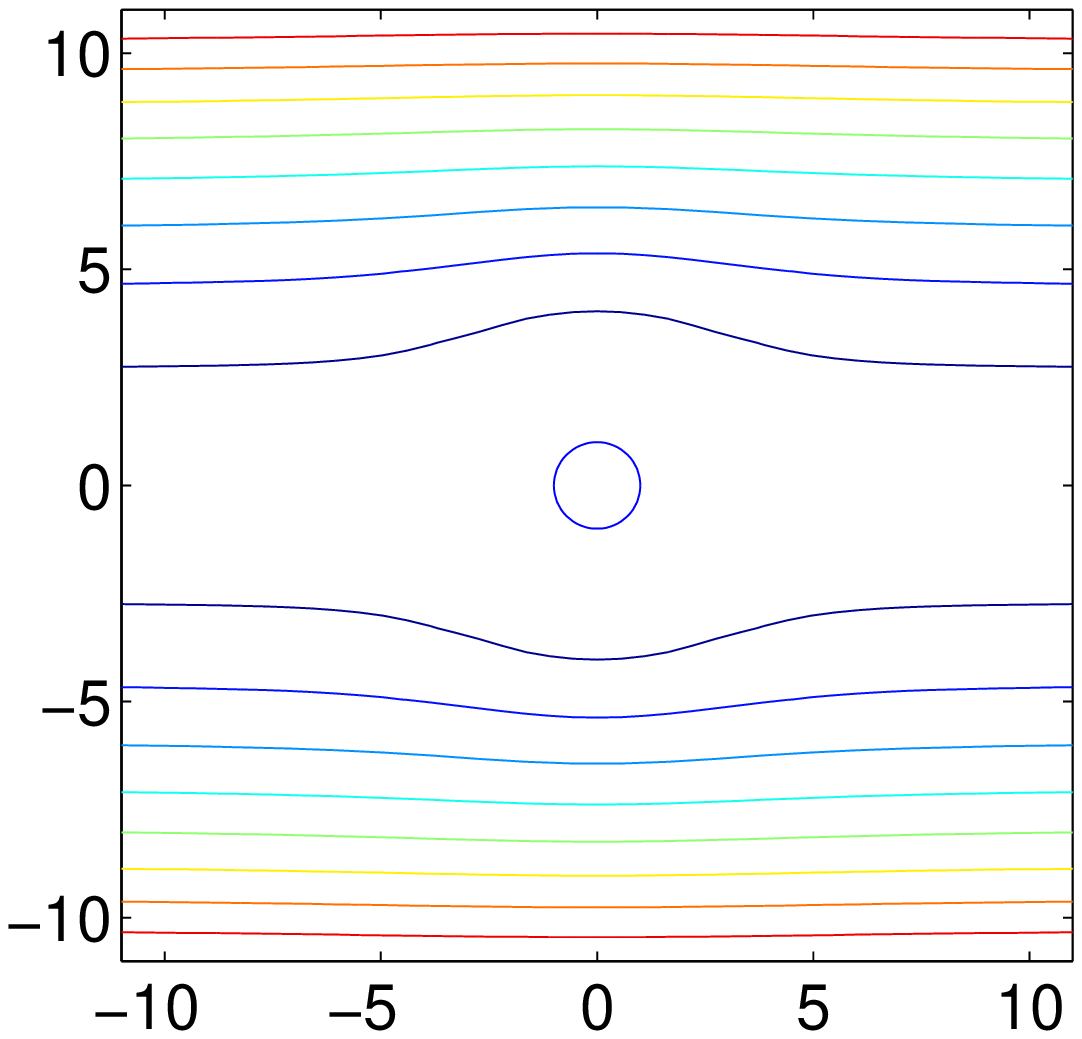}}
\hskip -0.3in
\subfigure[]{\includegraphics[height=1.3in]{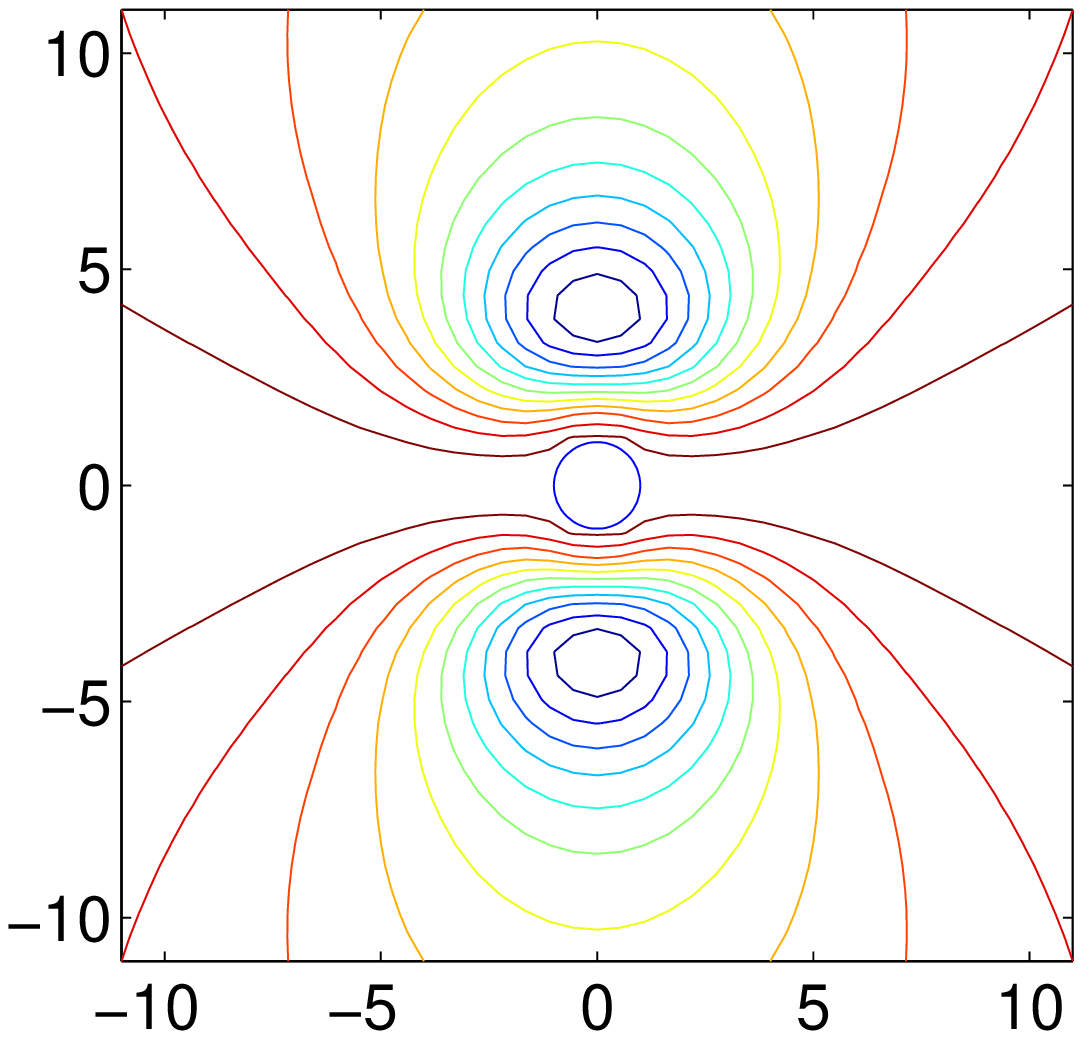}}
\hskip -0.3in
\subfigure[]{\includegraphics[height=1.3in]{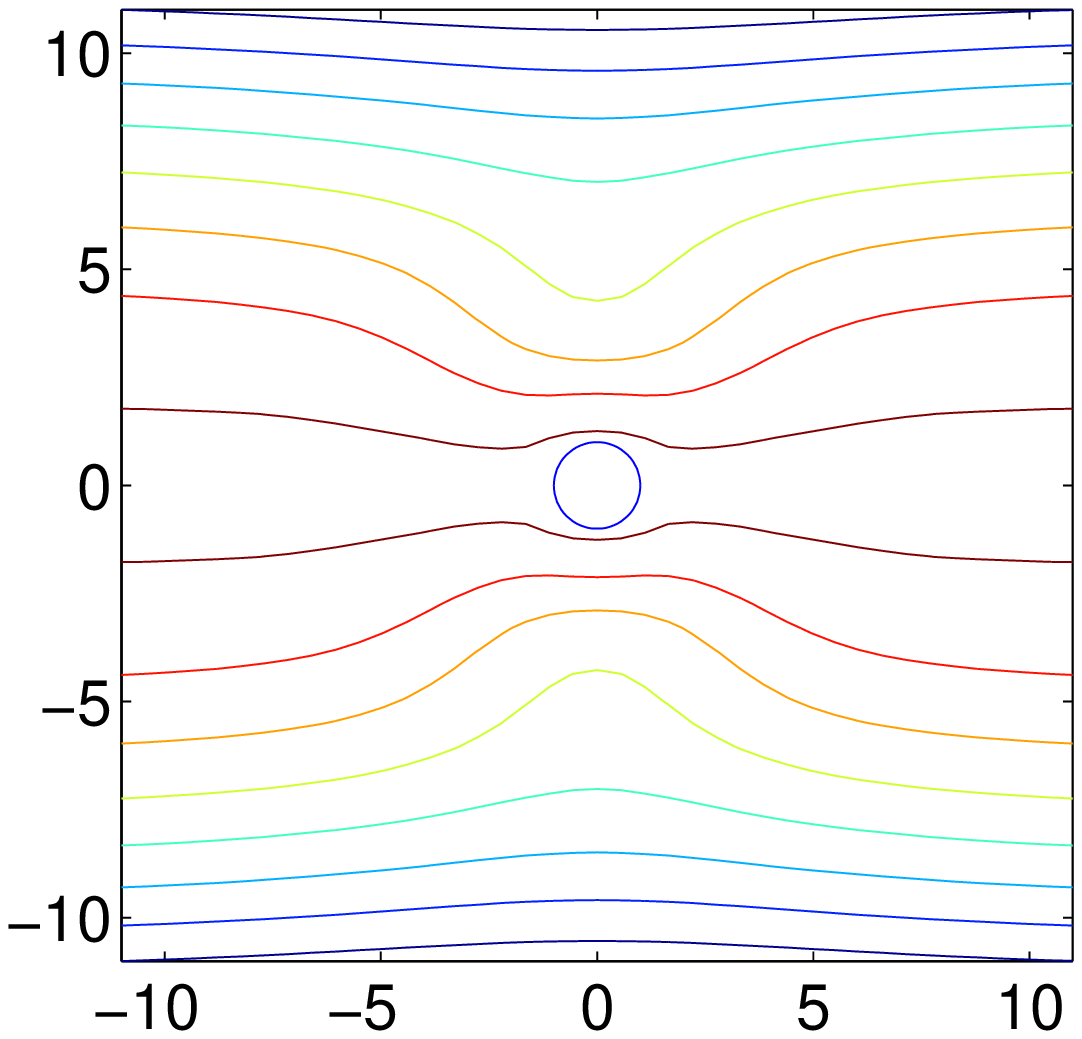}}
\caption[]{Inertial Stokes with $G(t)=\delta(t)$, yielding frequency parameter $\beta=1$ and Weissenberg number $\Wi=0$. Times $\omega t=0, 0.45\pi, 0.5\pi, 0.51\pi$ are plotted.}
\label{fig:Unsteady_Stokes}
\end{figure}

\begin{figure}
\centering
\hskip -0.2in
\subfigure[]{\includegraphics[height=1.2in]{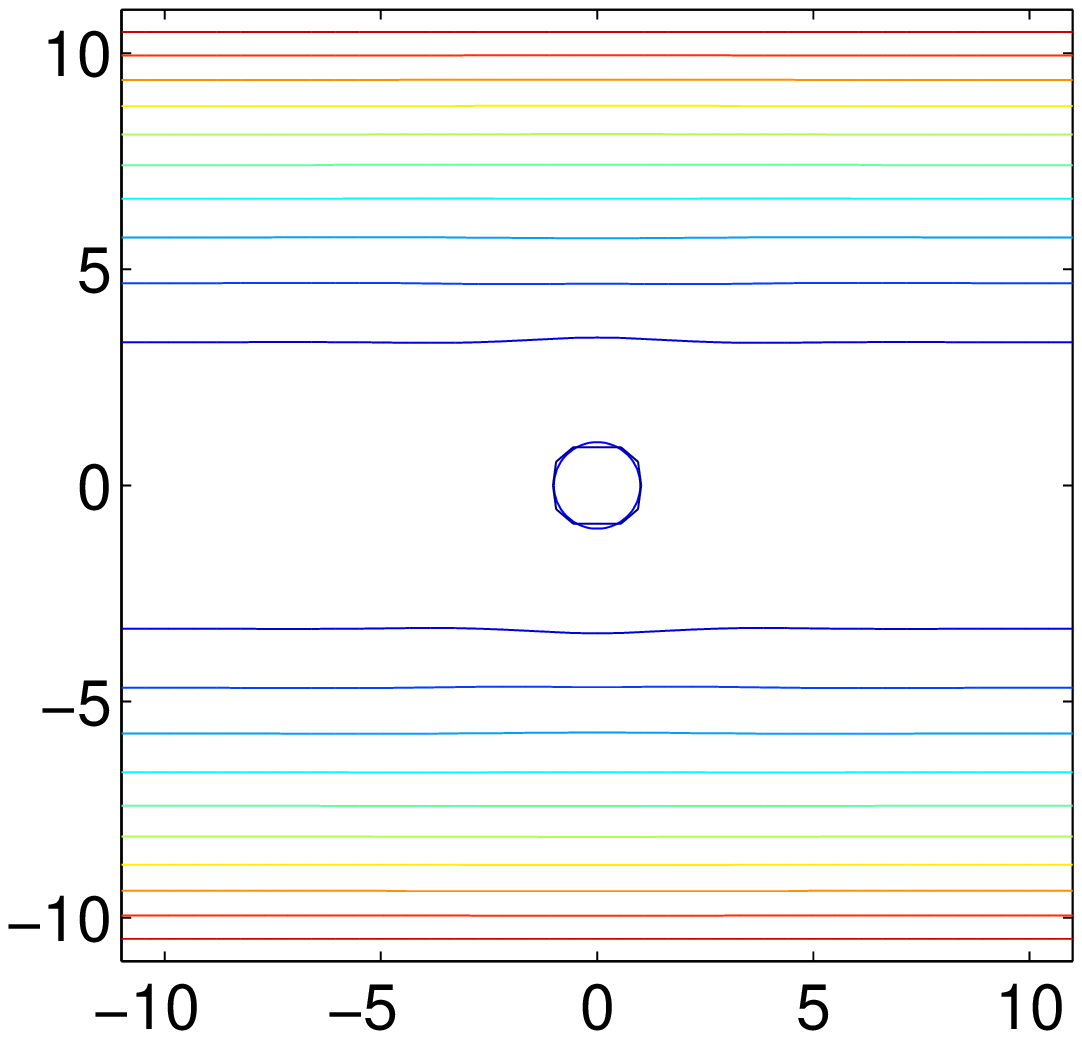}}
\hskip -0.3in
\subfigure[]{\includegraphics[height=1.2in]{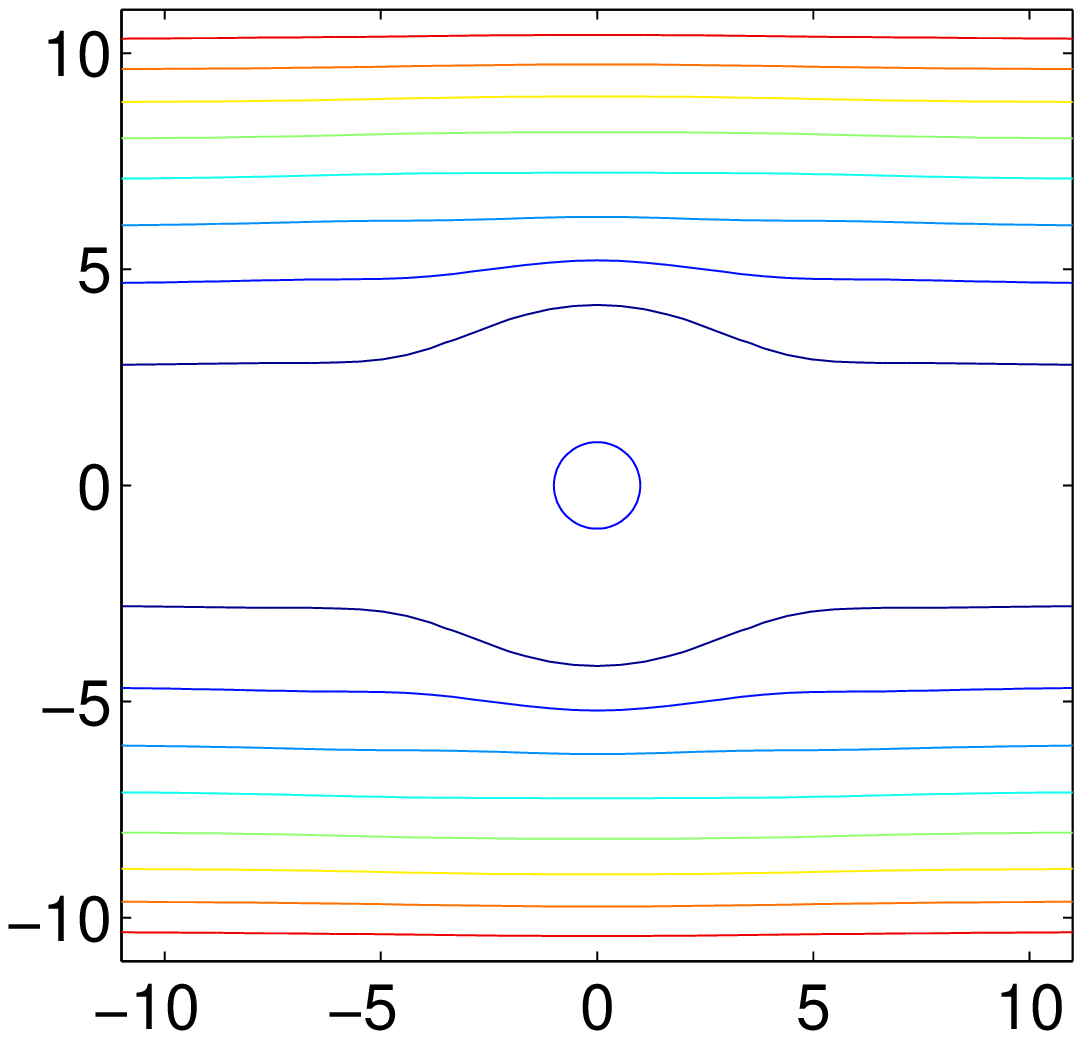}}
\hskip -0.3in
\subfigure[]{\includegraphics[height=1.2in]{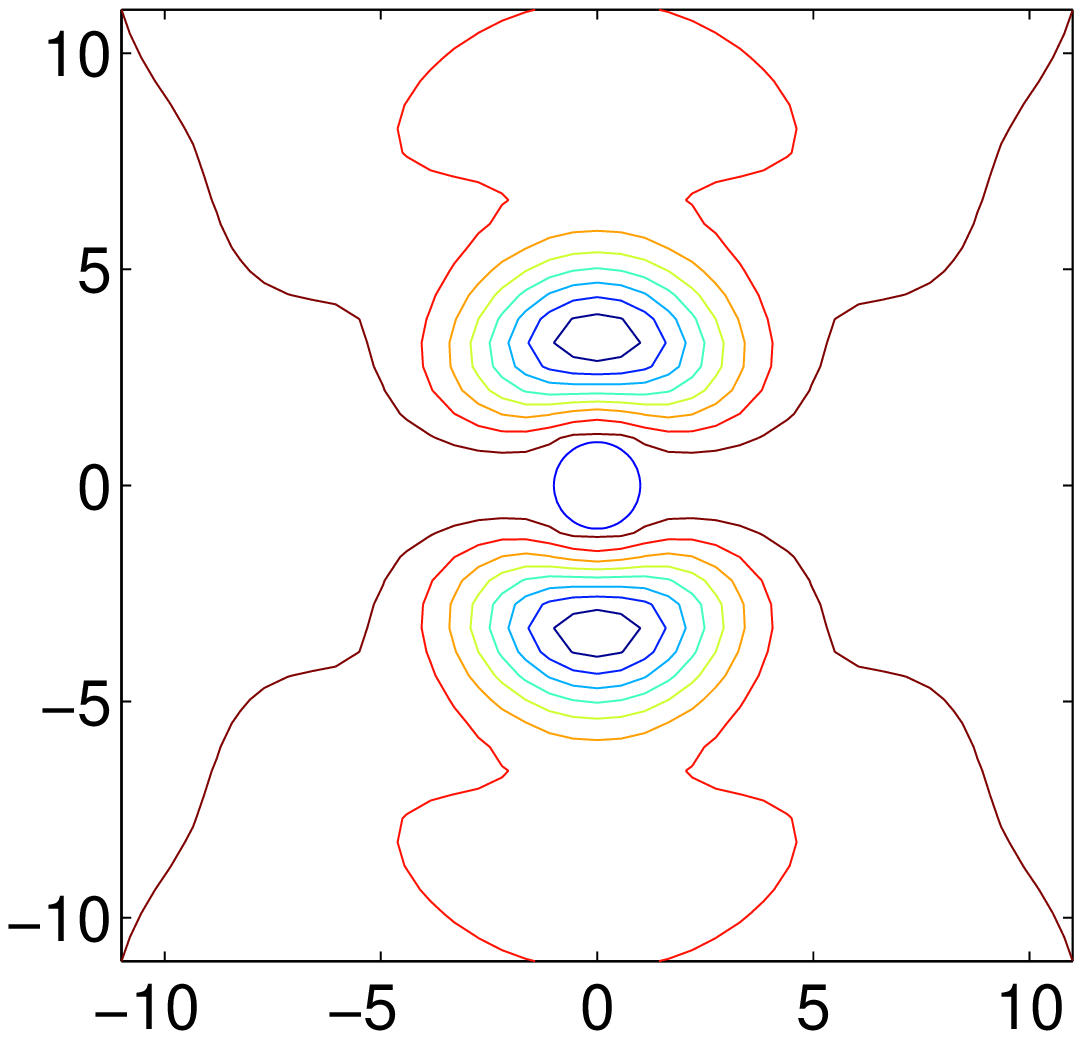}}
\hskip -0.3in
\subfigure[]{\includegraphics[height=1.2in]{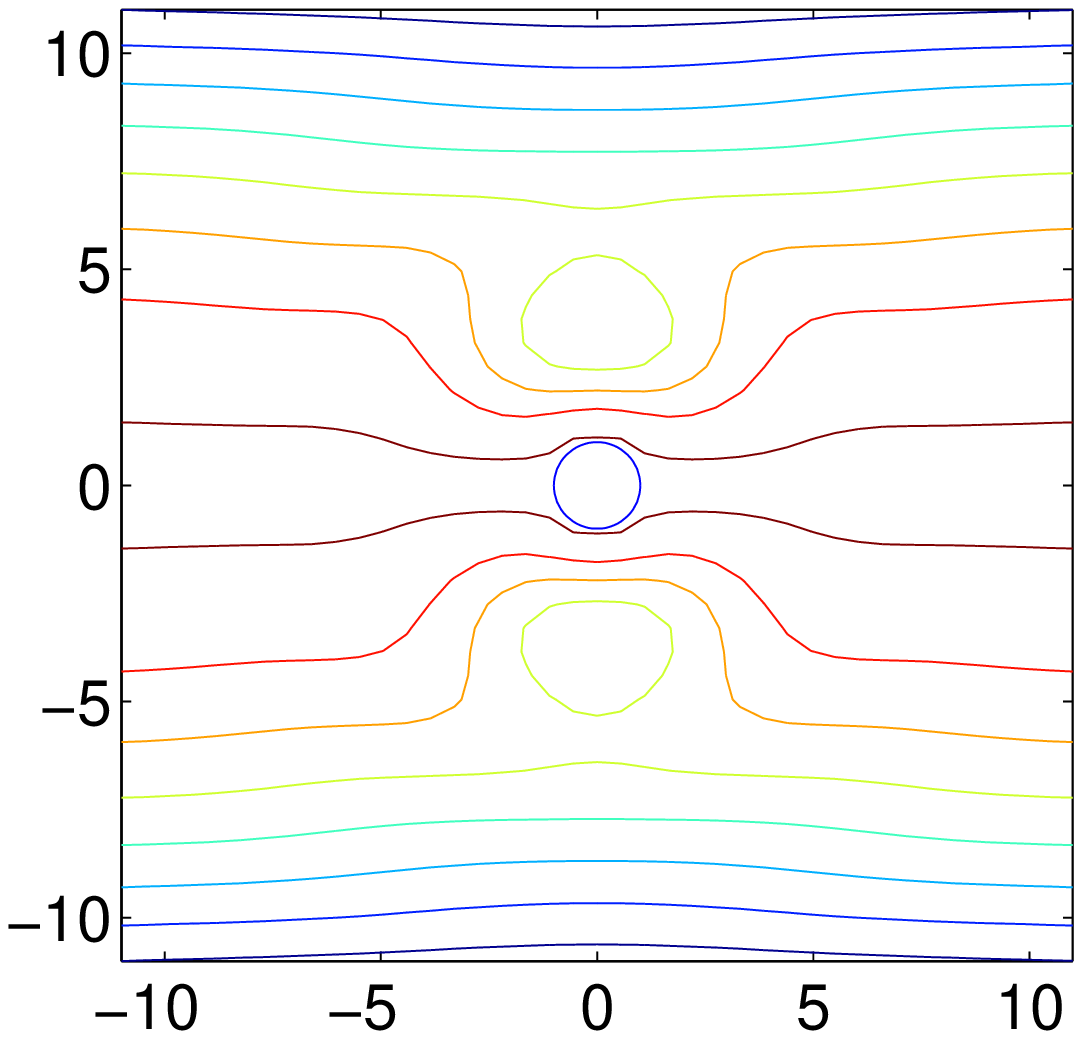}}
\caption[]{Inertial LVE with $G(t)=e^{-t}$, yielding frequency parameter $\beta=\sqrt{2}$ and Weissenberg number $\Wi=1.0$. Times $\omega t=0, 0.45\pi, 0.5\pi, 0.51\pi$ are plotted.}
\label{fig:LVE}
\end{figure}

As studied by Pozrikidis \citet{poz89}, when a spherical particle undergoes oscillatory motion in unsteady Stokes flow, there is an eddy generated on the particle, expanding and propagating into the fluid, during the deceleration stage within each period, see Fig. \ref{fig:Unsteady_Stokes}. With LVE, there are a few differences, see Fig. \ref{fig:LVE}. First, there is a sequence of eddies instead of one, with diminishing strengths. Second, the eddies are generated in the interior of the fluid and they barely travel. Third, the eddies are deferred in that they disappear during the acceleration phase -- recall that in the Newtonian case the eddy disappears exactly when the particle starts accelerating -- due to the elastic memory effect in LVE. These differences should bear consequences in engineering applications, e.g., material dissolution and deposition.

\begin{figure}
\centering
\hskip -0.2in
\subfigure[]{\includegraphics[height=1.2in]{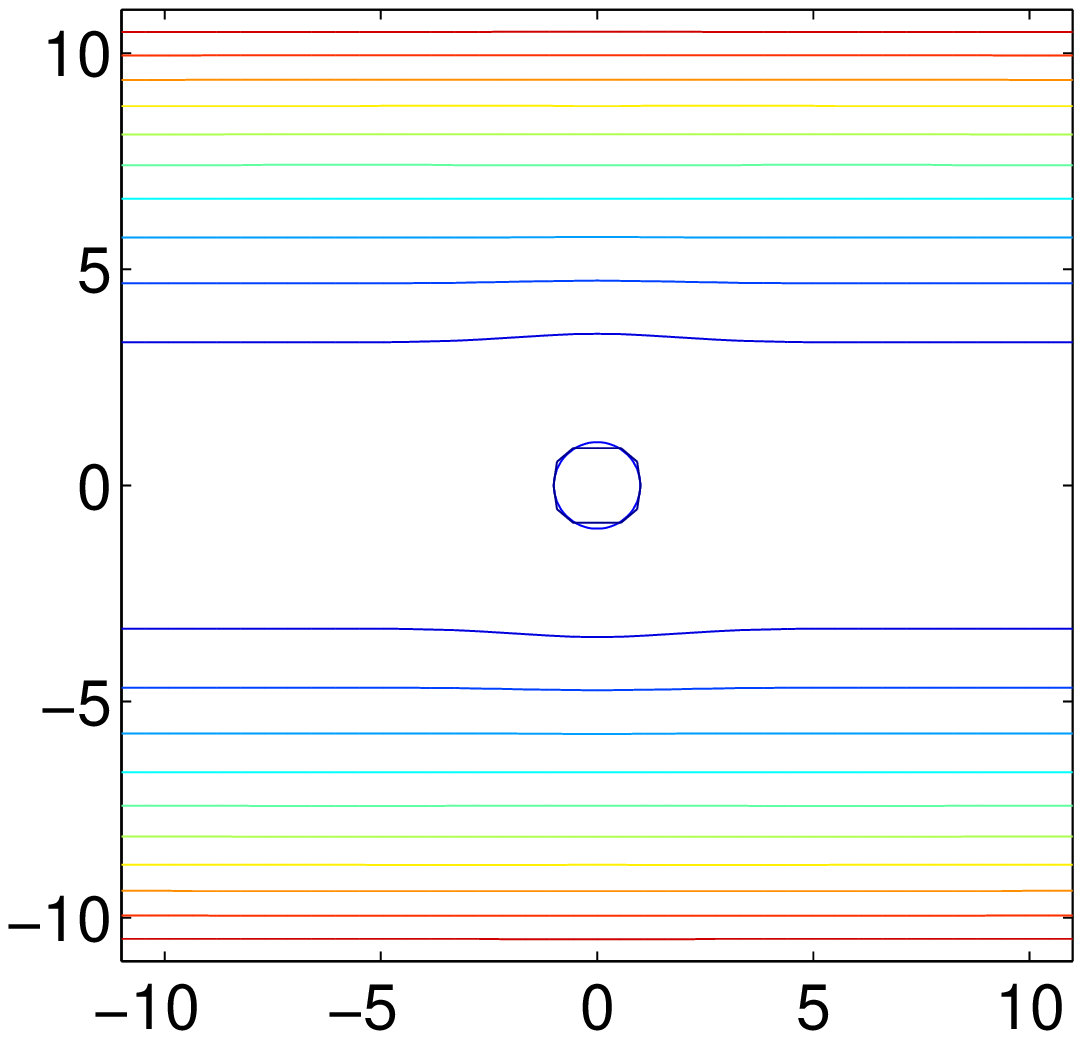}}
\hskip -0.3in
\subfigure[]{\includegraphics[height=1.2in]{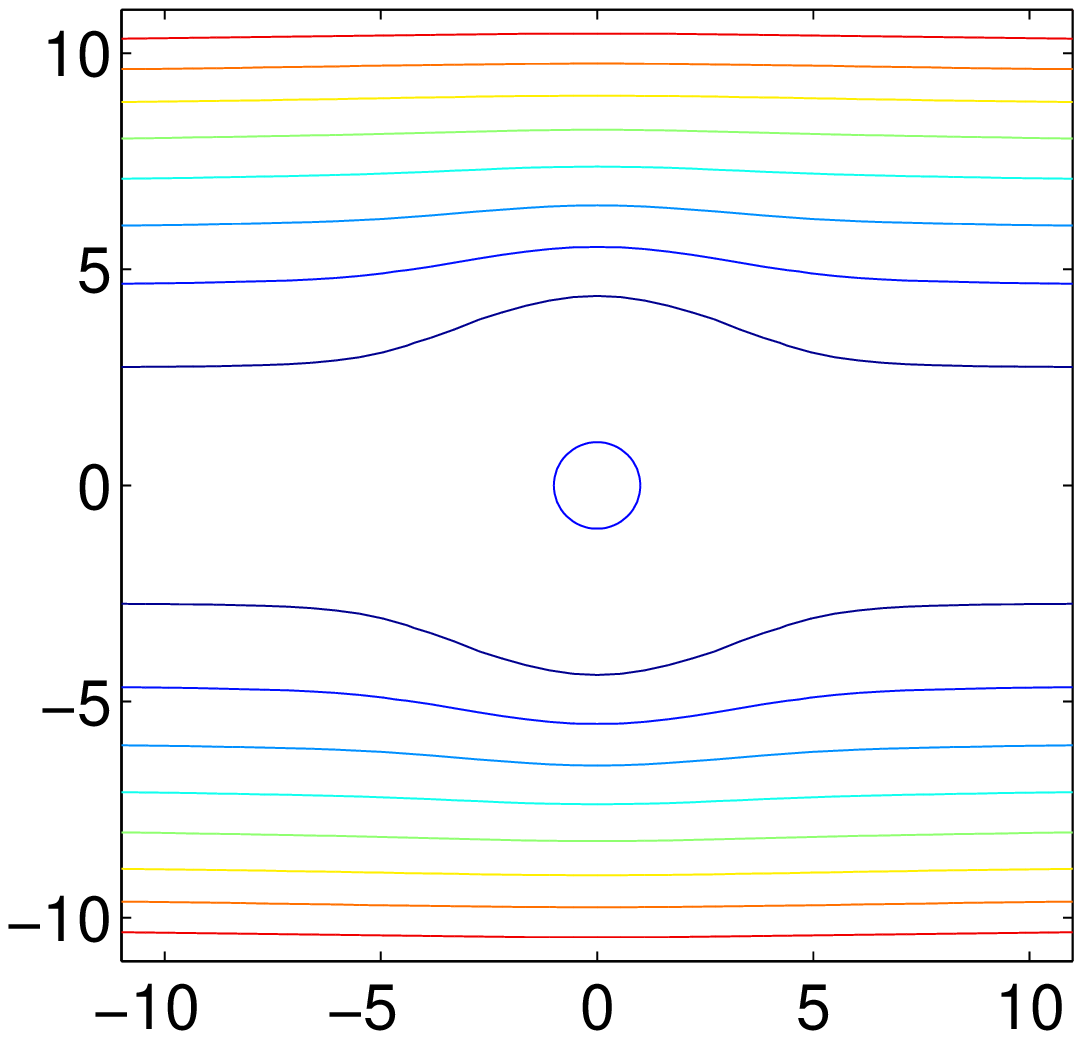}}
\hskip -0.3in
\subfigure[]{\includegraphics[height=1.2in]{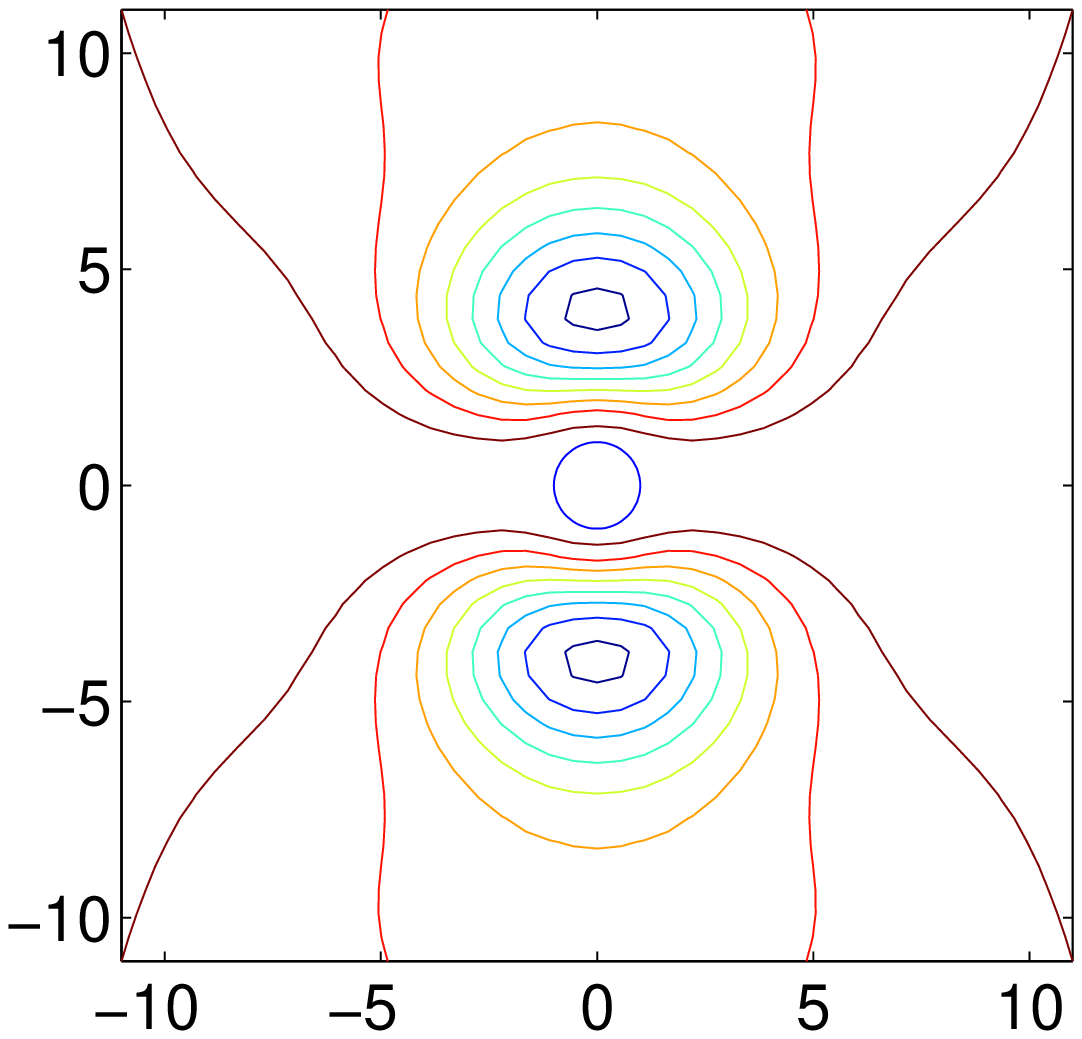}}
\hskip -0.3in
\subfigure[]{\includegraphics[height=1.2in]{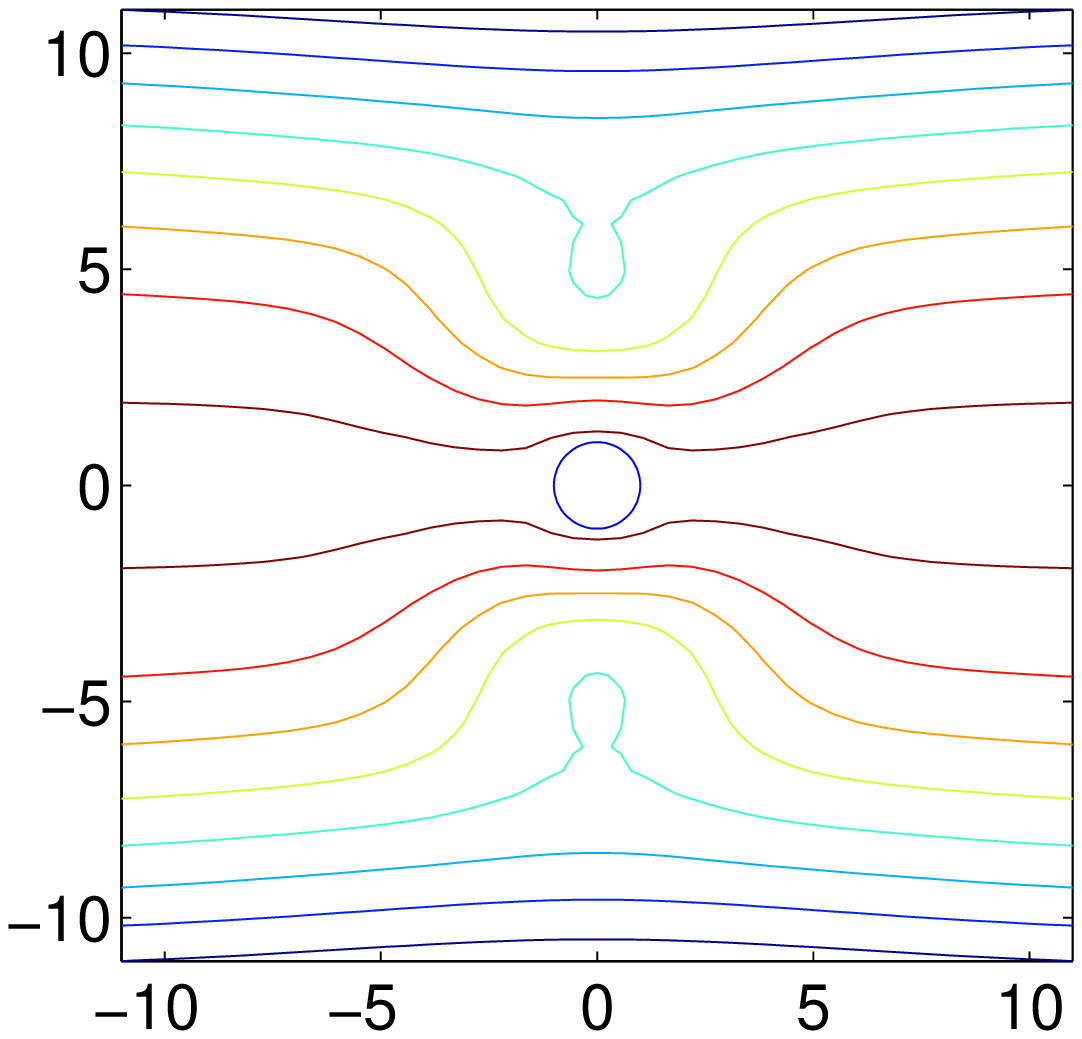}}
\caption[]{Inertial Newtonian+LVE with $G(t)=0.5\delta(t)+e^{-t}$, yielding frequency parameter $\beta=\sqrt{2/5}$ and Weissenberg number $\Wi=1.0$. Times $\omega t=0, 0.45\pi, 0.5\pi, 0.51\pi$ are plotted.}
\label{fig:New_LVE}
\end{figure}

We make the following comments. First, we also plotted the streamlines for the the case of a fluid with both a Newtonian component and an LVE component, with relaxation modulus $G(t)=0.5\delta(t)+e^{-t}$, see Fig. \ref{fig:New_LVE}. One can see that the streamline pattern is between that for the Newtonian case and that for the pure LVE case. There is still a sequence of eddies, with all but the largest one barely visible.  Since the existence of a Newtonian component does not qualitatively alter the flow pattern, from now on we will concentrate on studying pure LVE fluid with no Newtonian component. Second, we note that LVE with multiple modes give rise to similar streamline patterns (not plotted), so we will only include one mode in our LVE relaxation modulus. Note that adding a Newtonian component or using multiple modes in LVE is straightforward and adds no computational cost.

\begin{figure}
\centering
\subfigure[]{\includegraphics[height=1.8in]{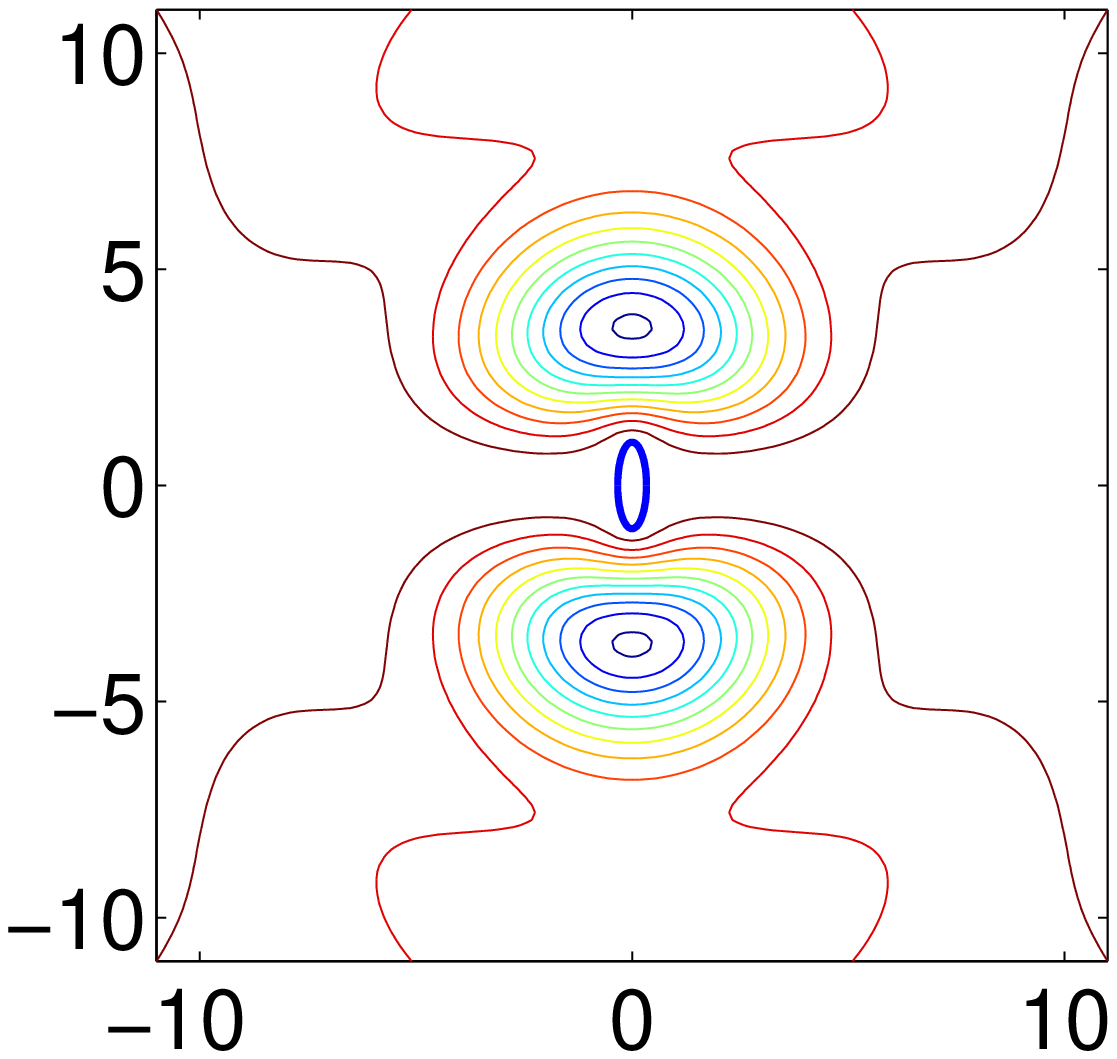}}
\subfigure[]{\includegraphics[height=1.8in]{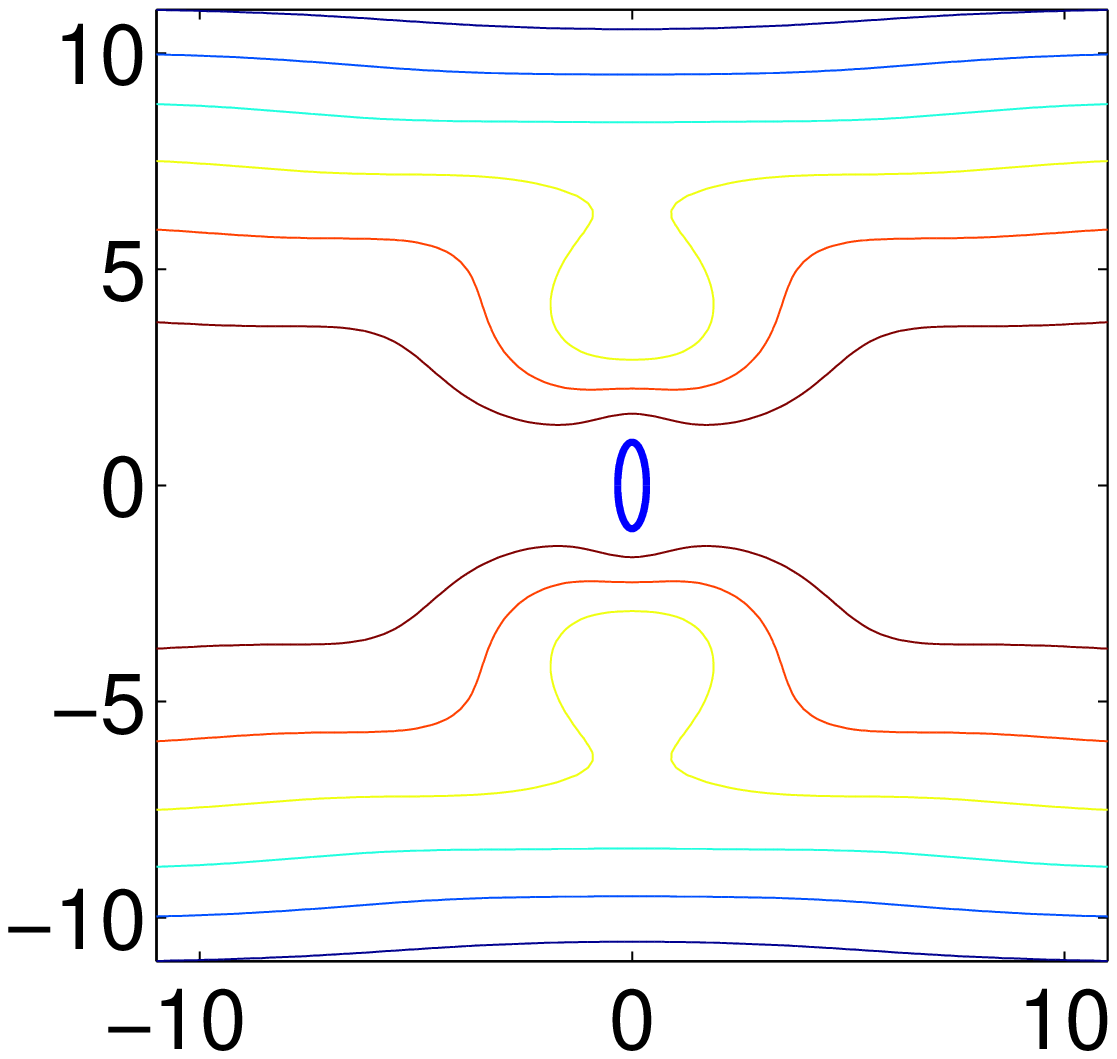}}
\caption[]{Oblate spheroid with frequency parameter $\beta=1.0$ and Weissenberg number $\Wi=1.0$. Times $\omega t=0.5\pi$ (a), and $0.51\pi $ (b) are plotted.}
\label{fig:ob}
\end{figure}

\begin{figure}
\centering
\subfigure[]{\includegraphics[height=1.8in]{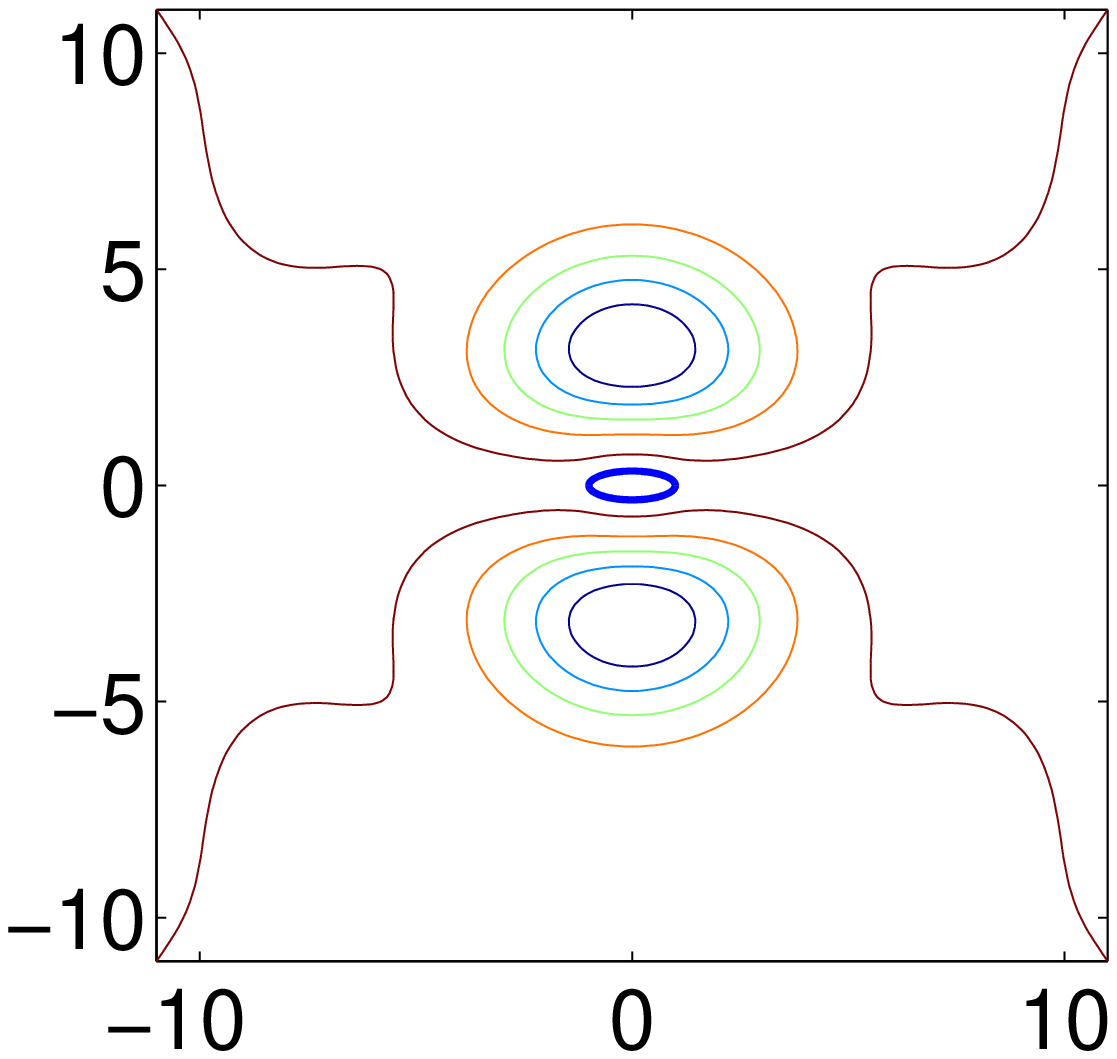}}
\subfigure[]{\includegraphics[height=1.8in]{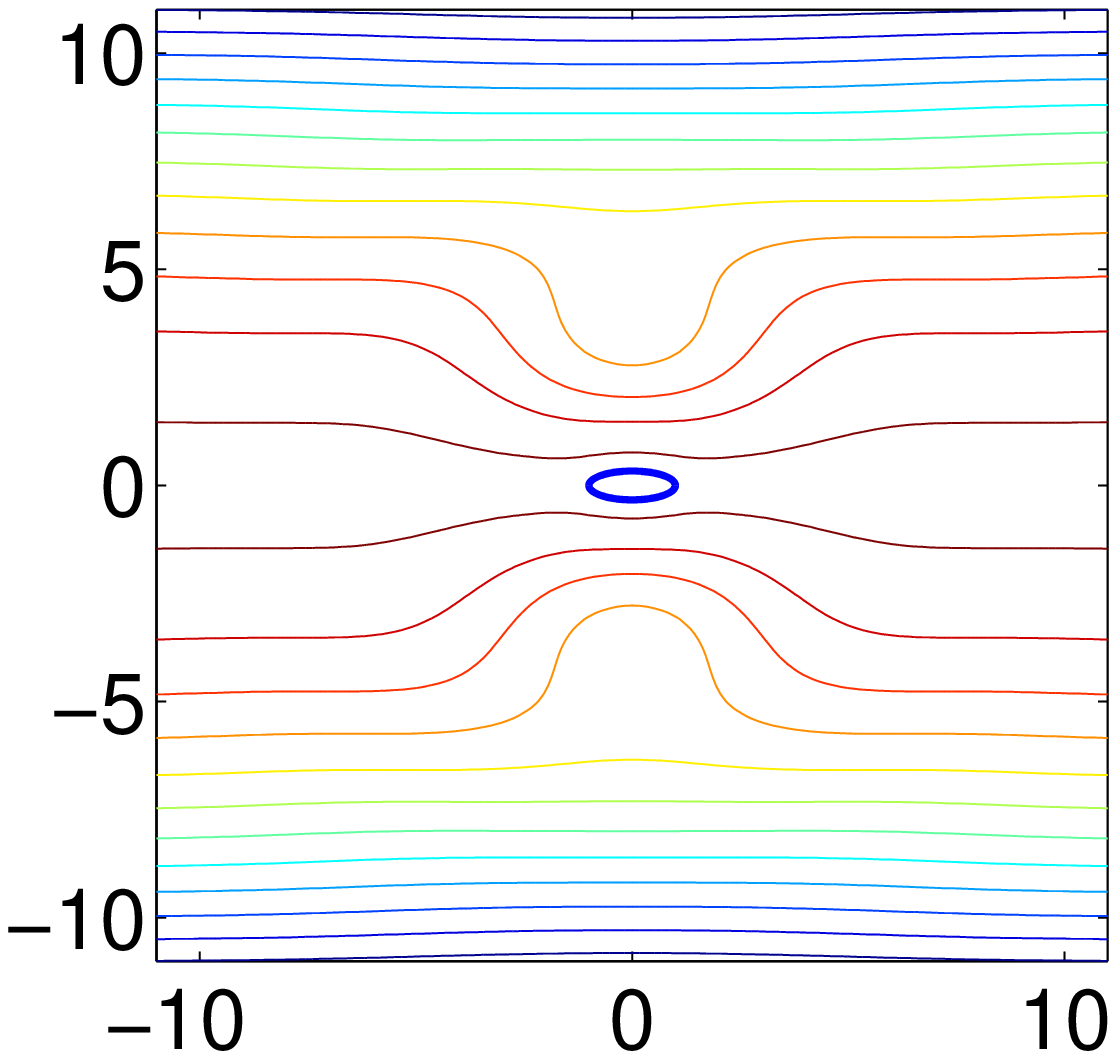}}
\caption[]{Prolate spheroid with frequency parameter $\beta=1.0$ and Weissenberg number $\Wi=1.0$. Times $\omega t=0.5\pi$ (a), and $0.51\pi $ (b) are plotted.}
\label{fig:pro}
\end{figure}

Analytic solutions exist for the case of a spherical particle, for both the Newtonian case \citep{Stokes_51} and the LVE case \citep{Schieber_13}, which was derived by using the correspondence principle between unsteady Stokes and LVE \citep{Zwanzig70}. Those solutions are used to check the accuracy of our method. Now we turn to the case of spheroids, for which there are no analytic solutions. We study two spheroids of aspect ratio 3:1, in prolate (long axis aligned with motion) and oblate (short axis aligned with motion) positions, and first plot the streamlines at characteristic times within one period. In Fig. \ref{fig:ob} and Fig. \ref{fig:pro}, it can be seen that the streamline patterns conform to the shape of the spheroids, and are similar to those in the case of a sphere.

\begin{figure}
\centering
\subfigure[]{\includegraphics[height=1.8in]{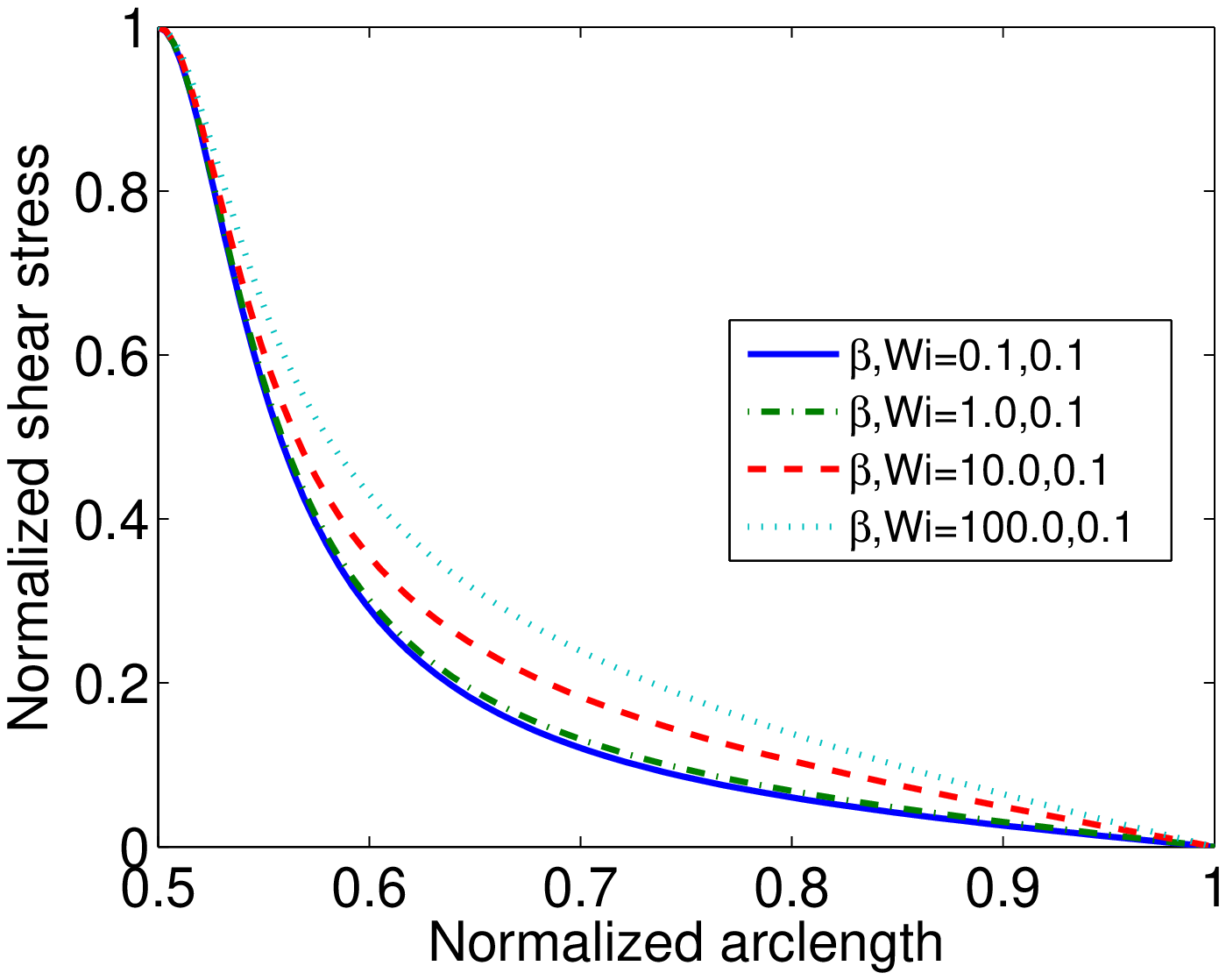}}
\subfigure[]{\includegraphics[height=1.8in]{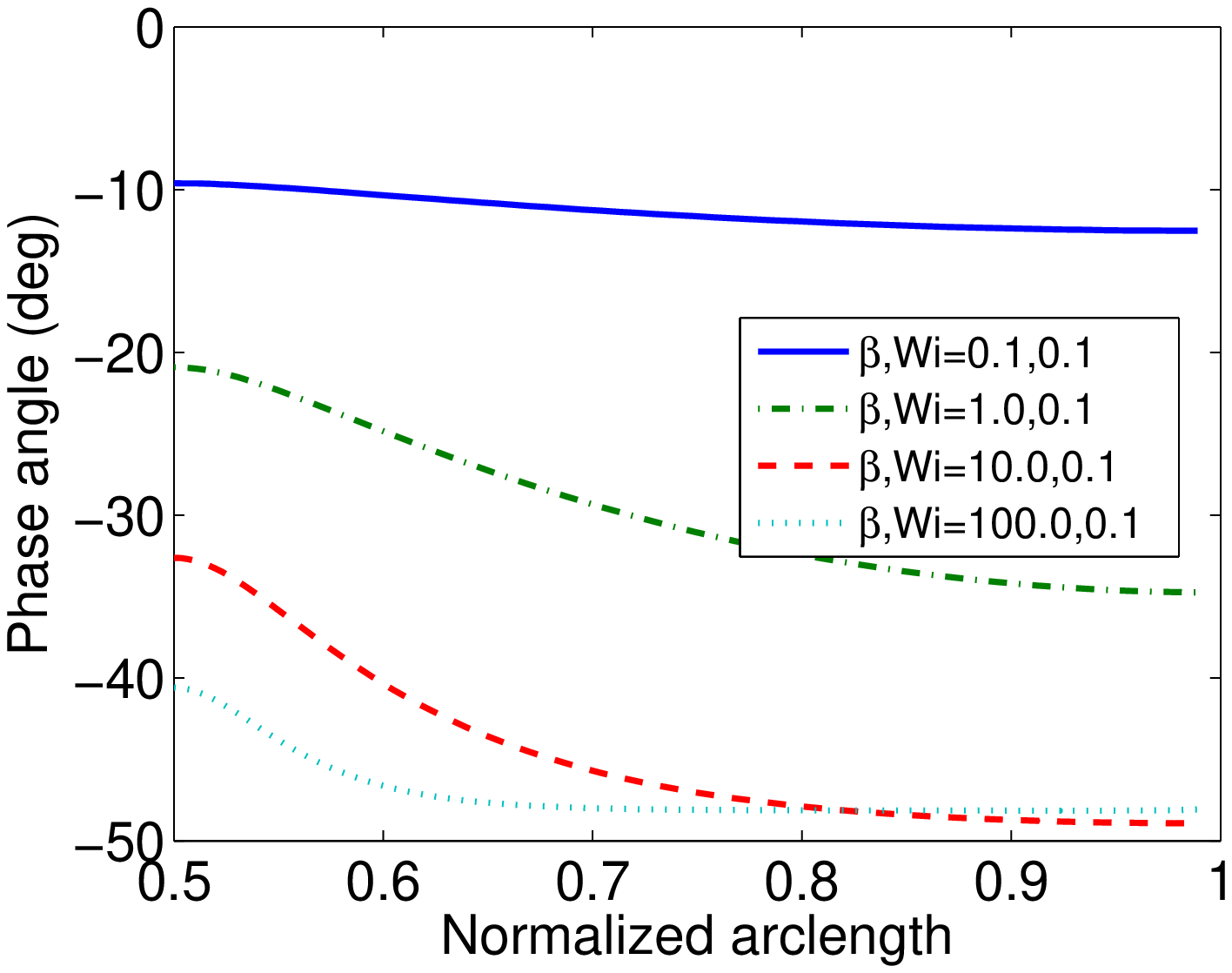}}
\hskip -0.4in
\subfigure[]{\includegraphics[height=1.8in]{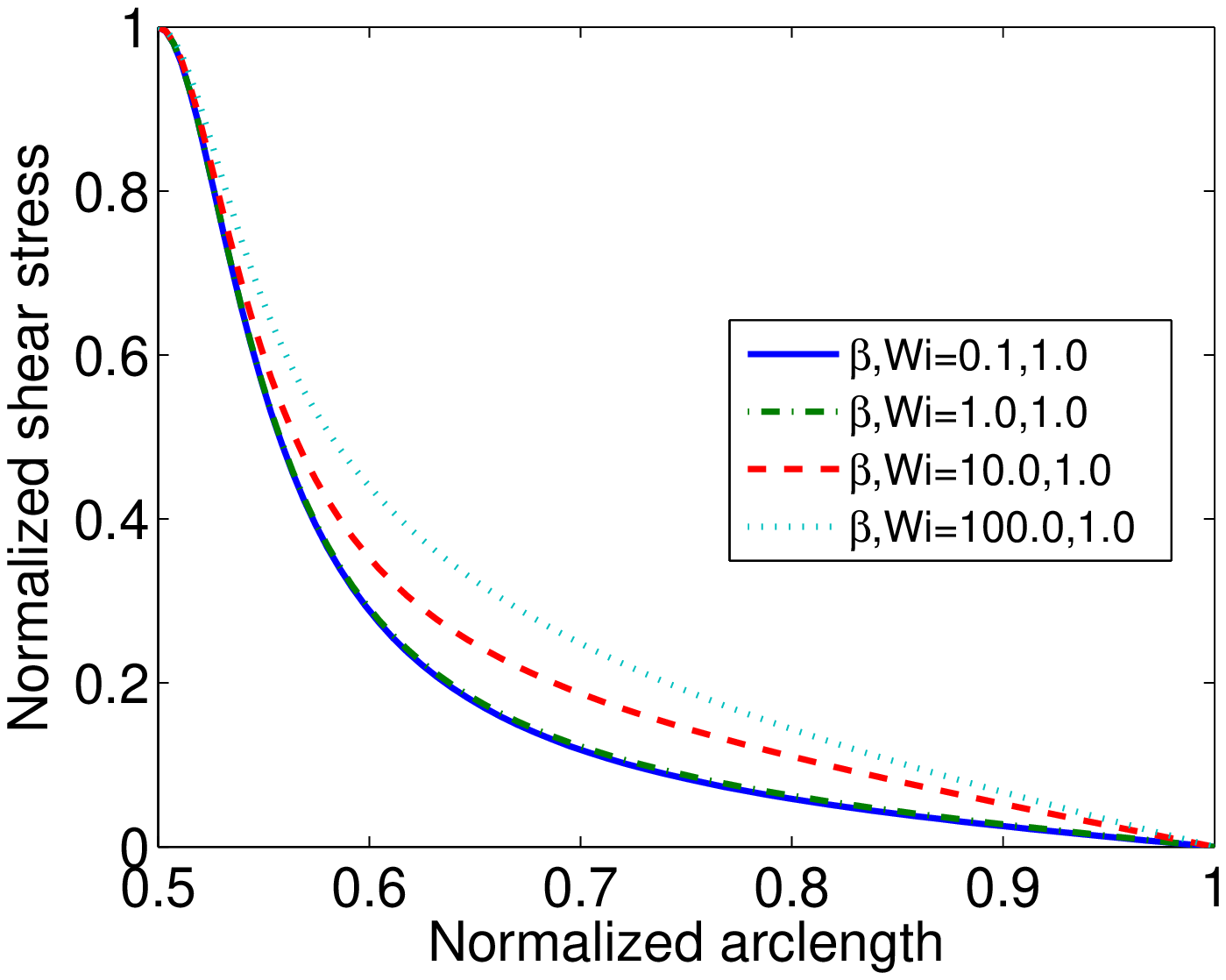}}
\subfigure[]{\includegraphics[height=1.8in]{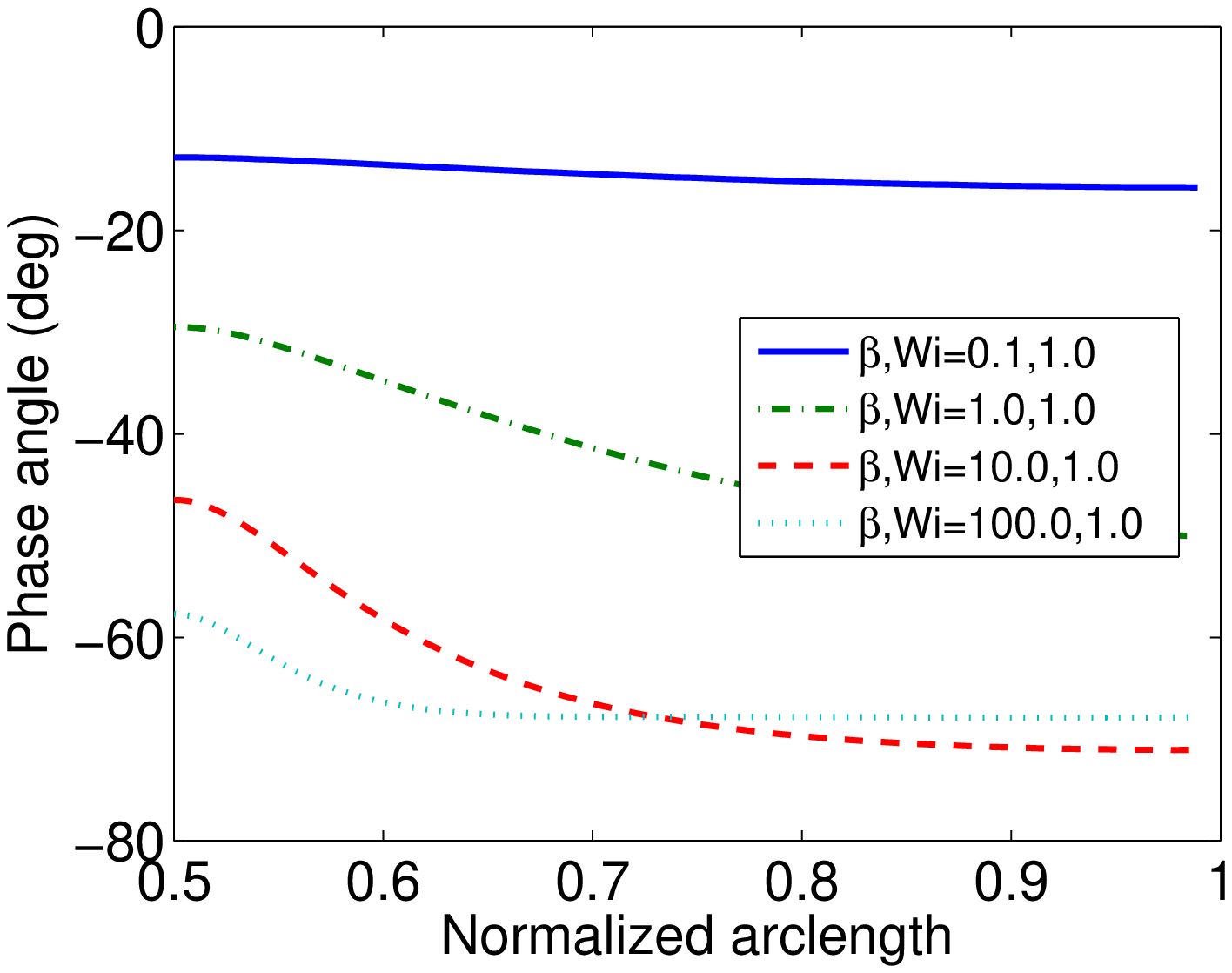}}
\hskip -0.4in
\subfigure[]{\includegraphics[height=1.8in]{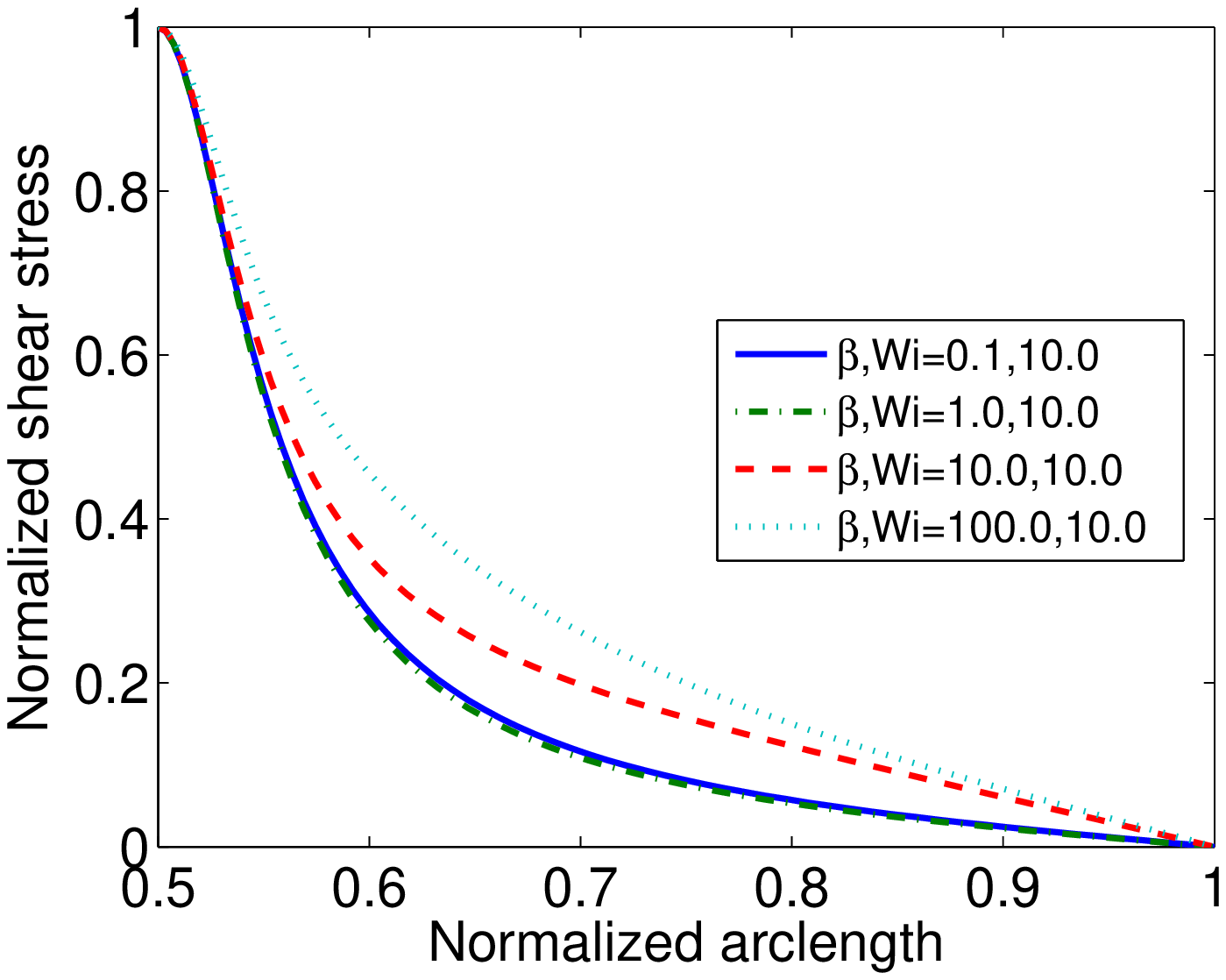}}
\subfigure[]{\includegraphics[height=1.8in]{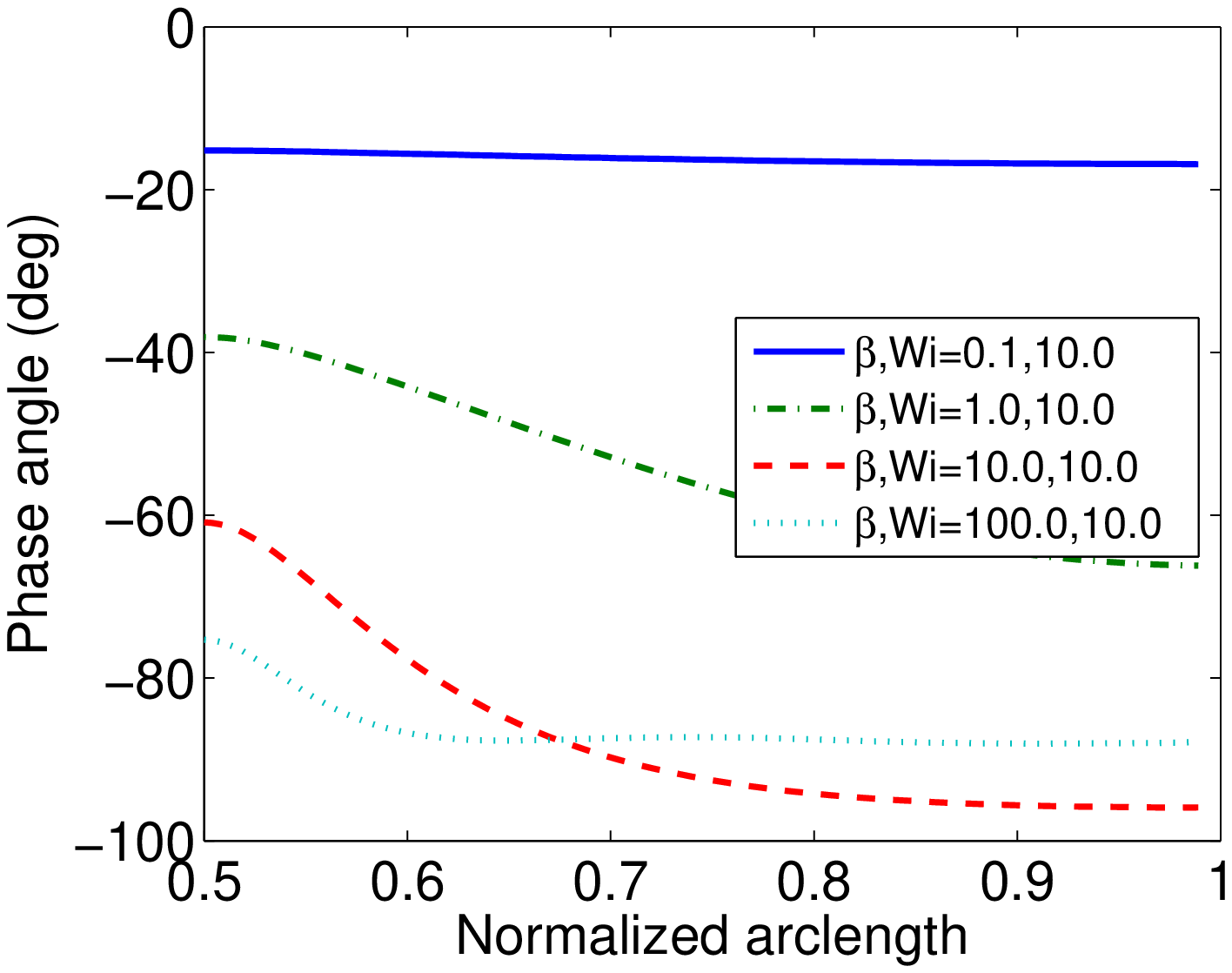}}
\hskip -0.4in
\subfigure[]{\includegraphics[height=1.8in]{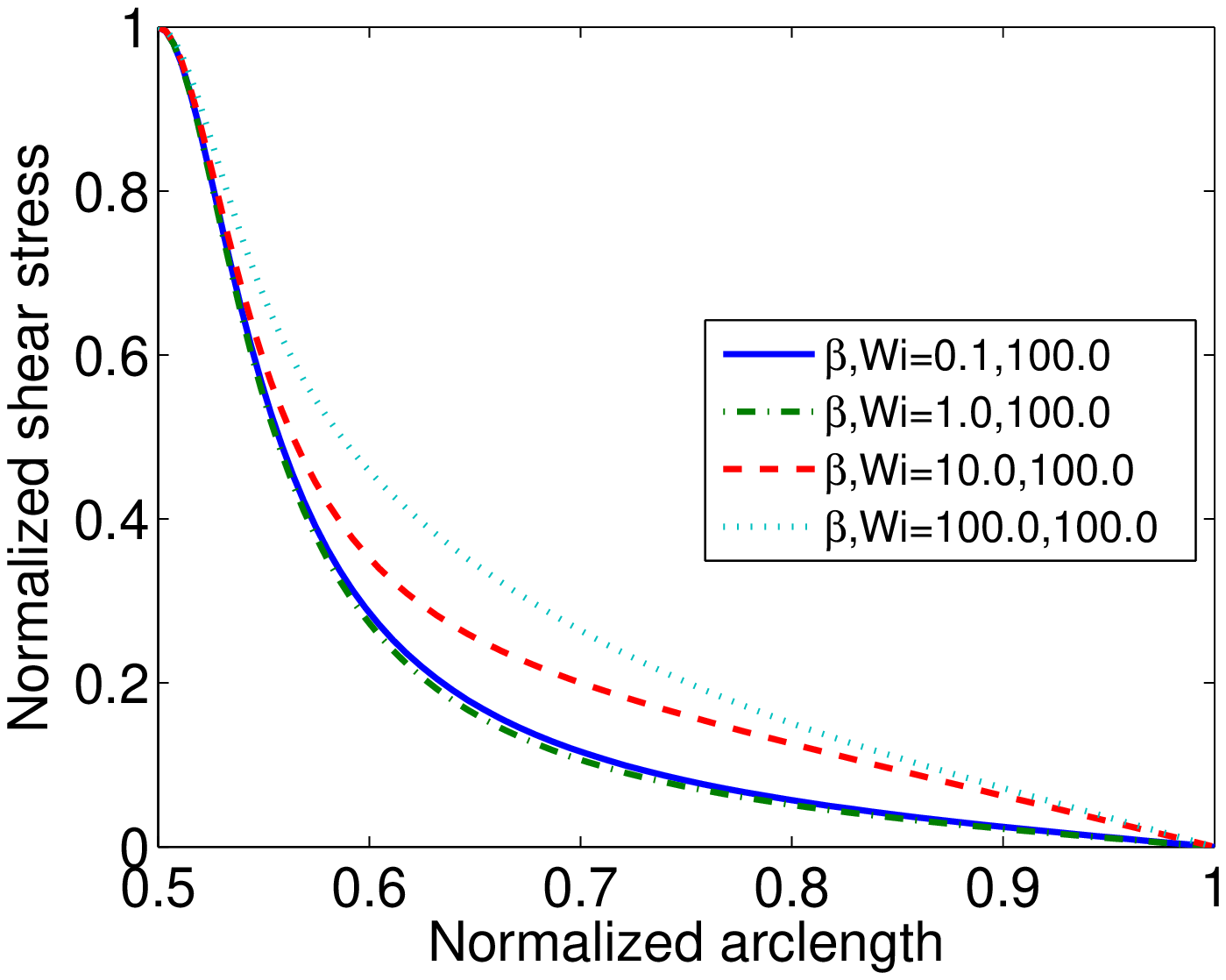}}
\subfigure[]{\includegraphics[height=1.8in]{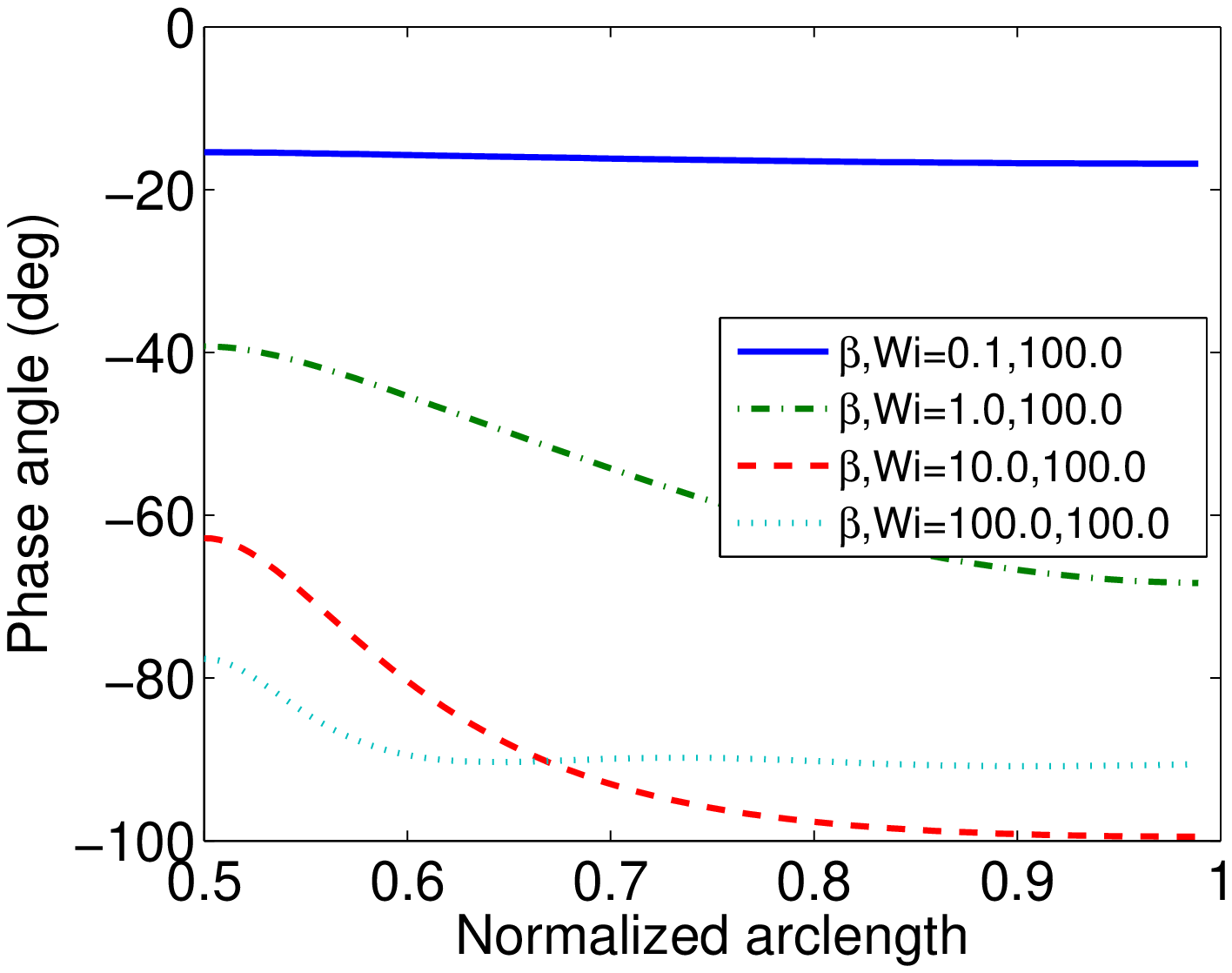}}
\caption[]{Amplitudes, normalized by its value at the midplane ($x=0$), and phases of shear stress on the oblate spheroid are plotted against the normalized arclength $s$, as the frequency parameter $\beta$ varies.}
\label{fig:obscn}
\end{figure}

\begin{figure}
\centering
\subfigure[]{\includegraphics[height=1.8in]{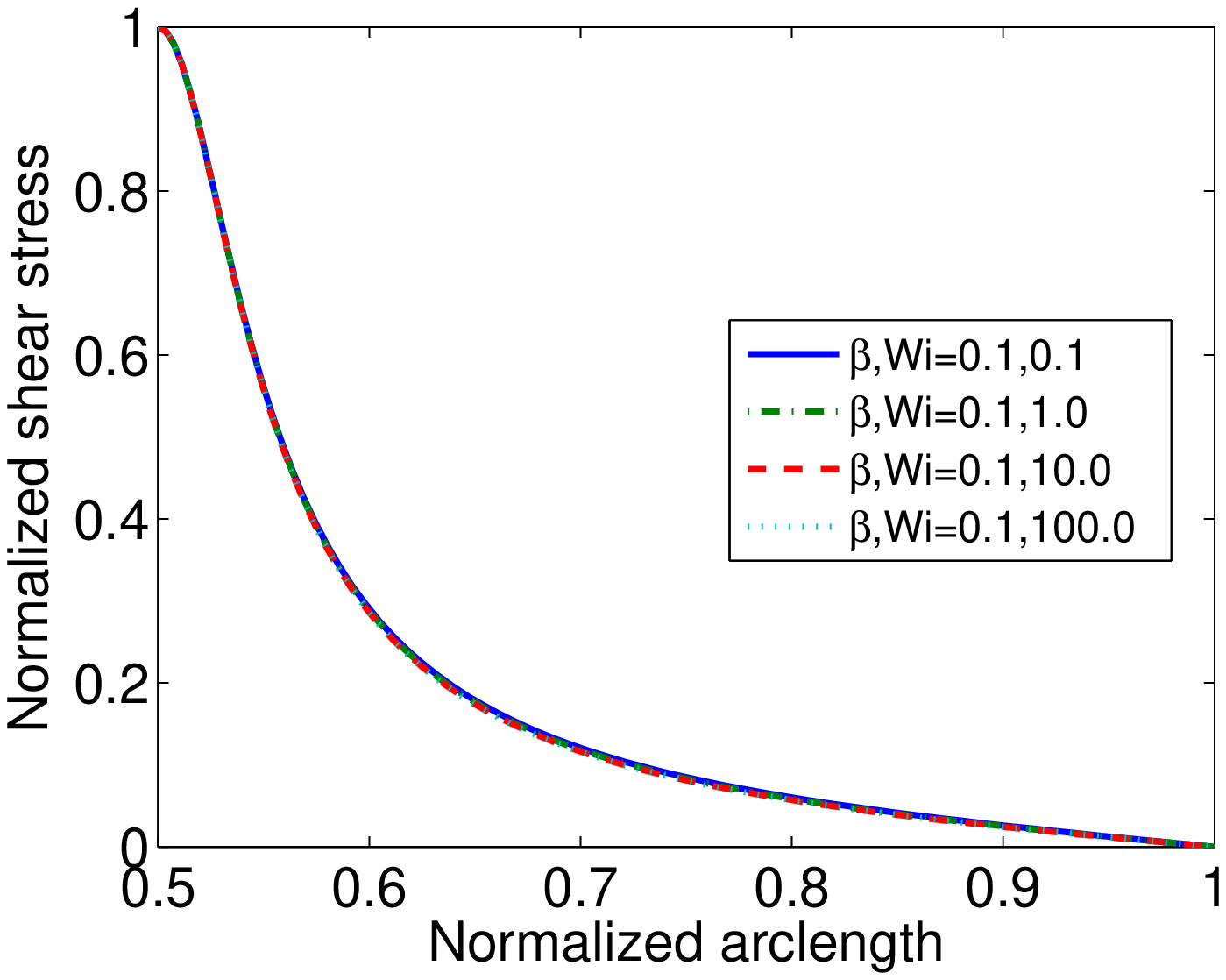}}
\subfigure[]{\includegraphics[height=1.8in]{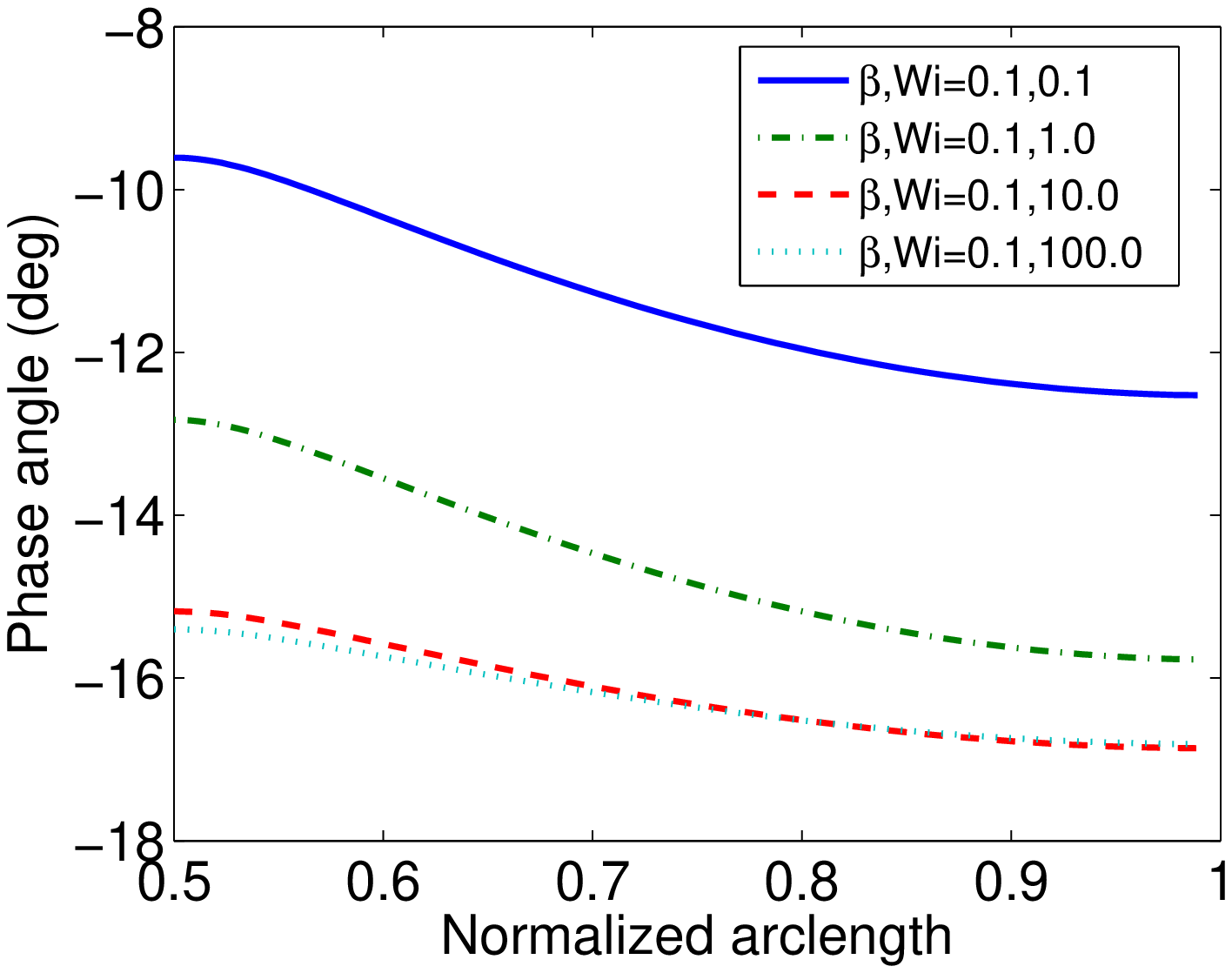}}
\hskip -0.4in
\subfigure[]{\includegraphics[height=1.8in]{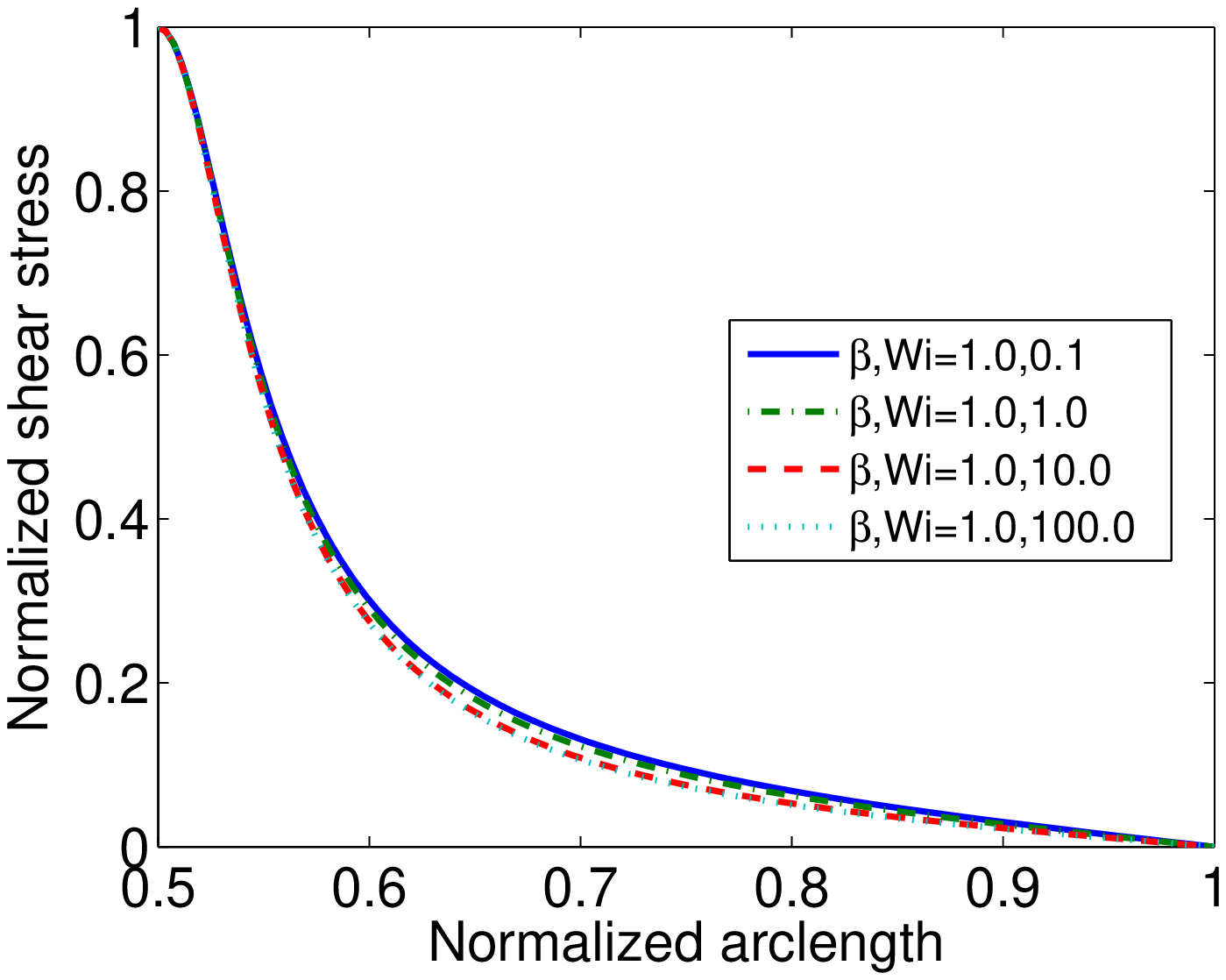}}
\subfigure[]{\includegraphics[height=1.8in]{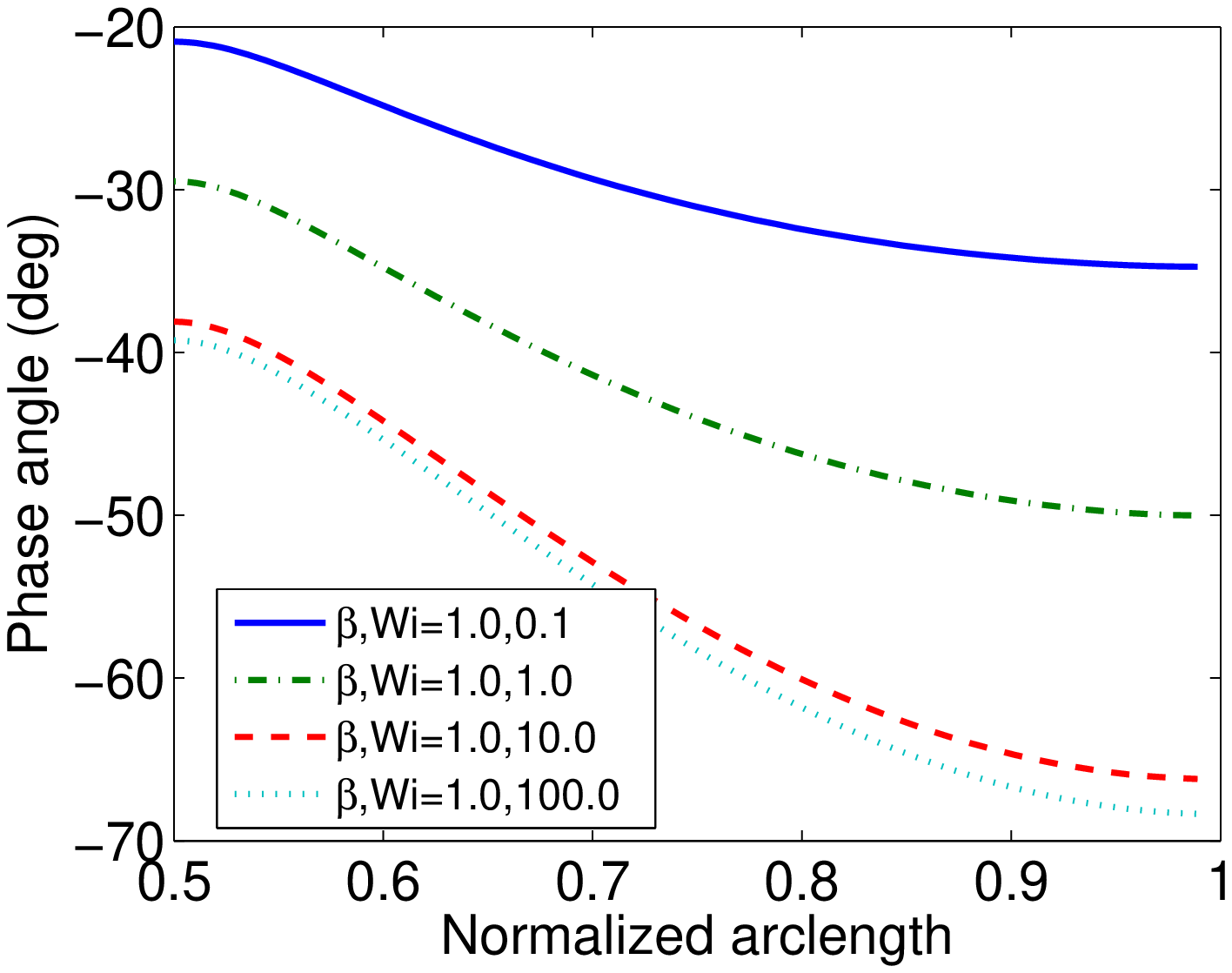}}
\hskip -0.4in
\subfigure[]{\includegraphics[height=1.8in]{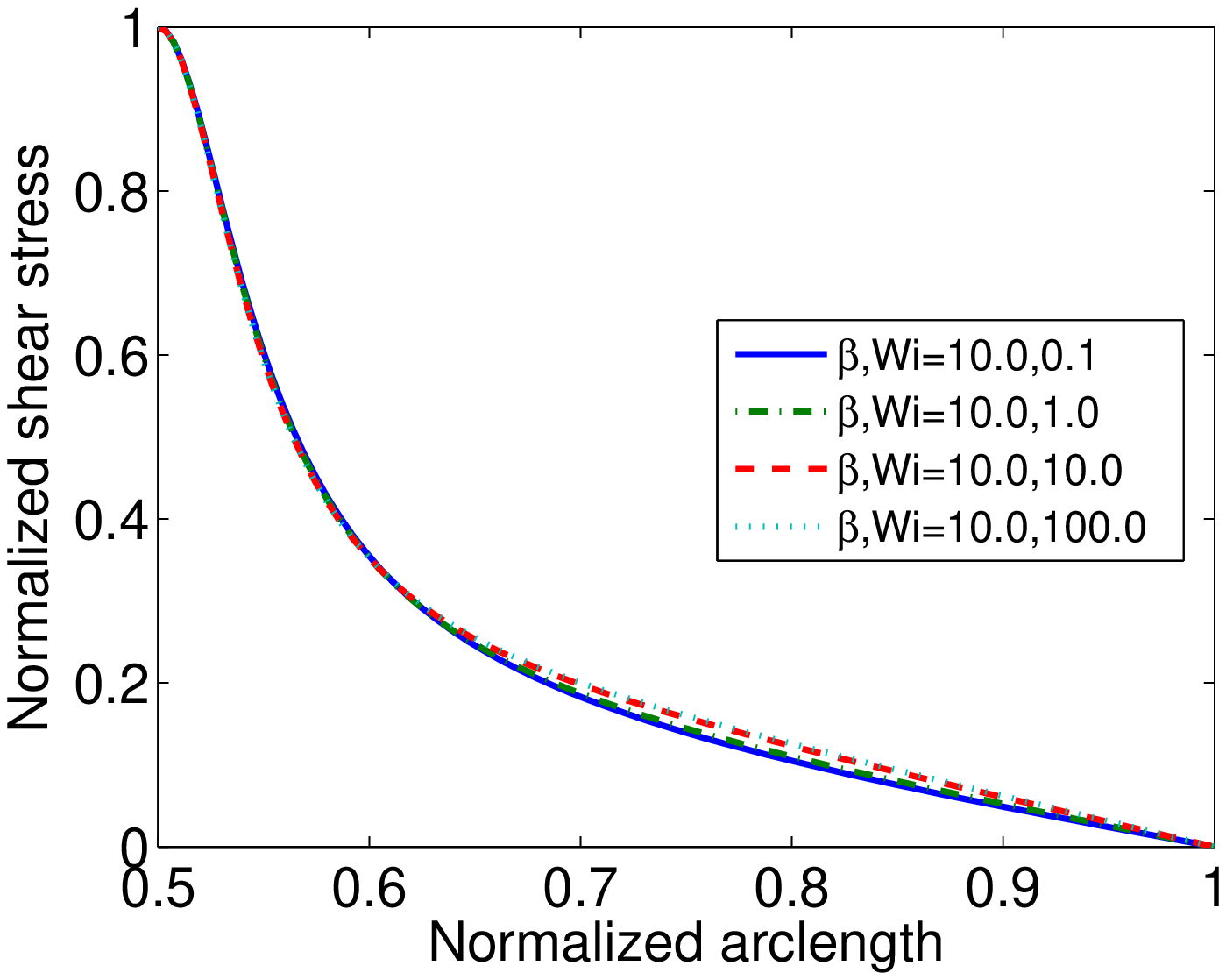}}
\subfigure[]{\includegraphics[height=1.8in]{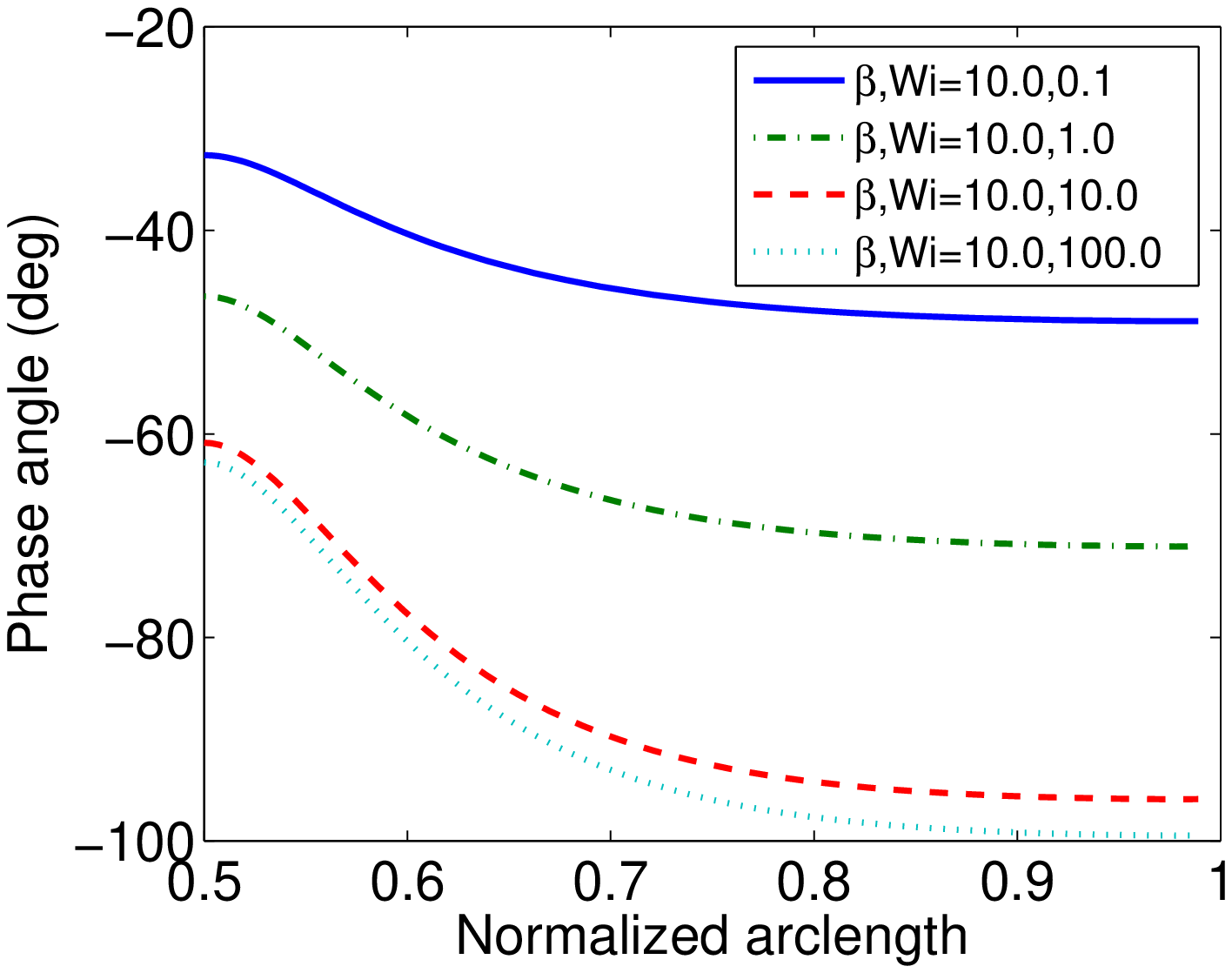}}
\hskip -0.4in
\subfigure[]{\includegraphics[height=1.8in]{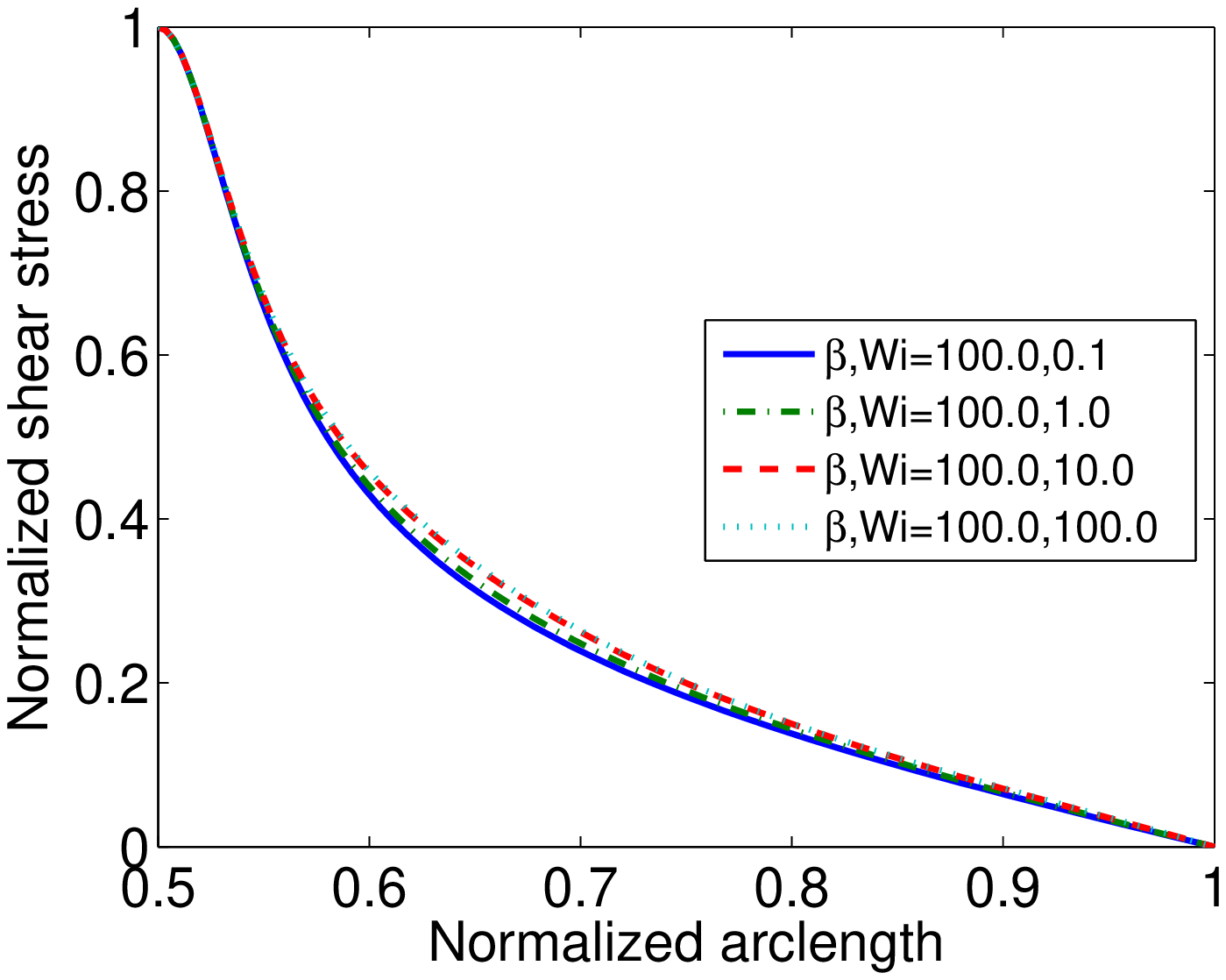}}
\subfigure[]{\includegraphics[height=1.8in]{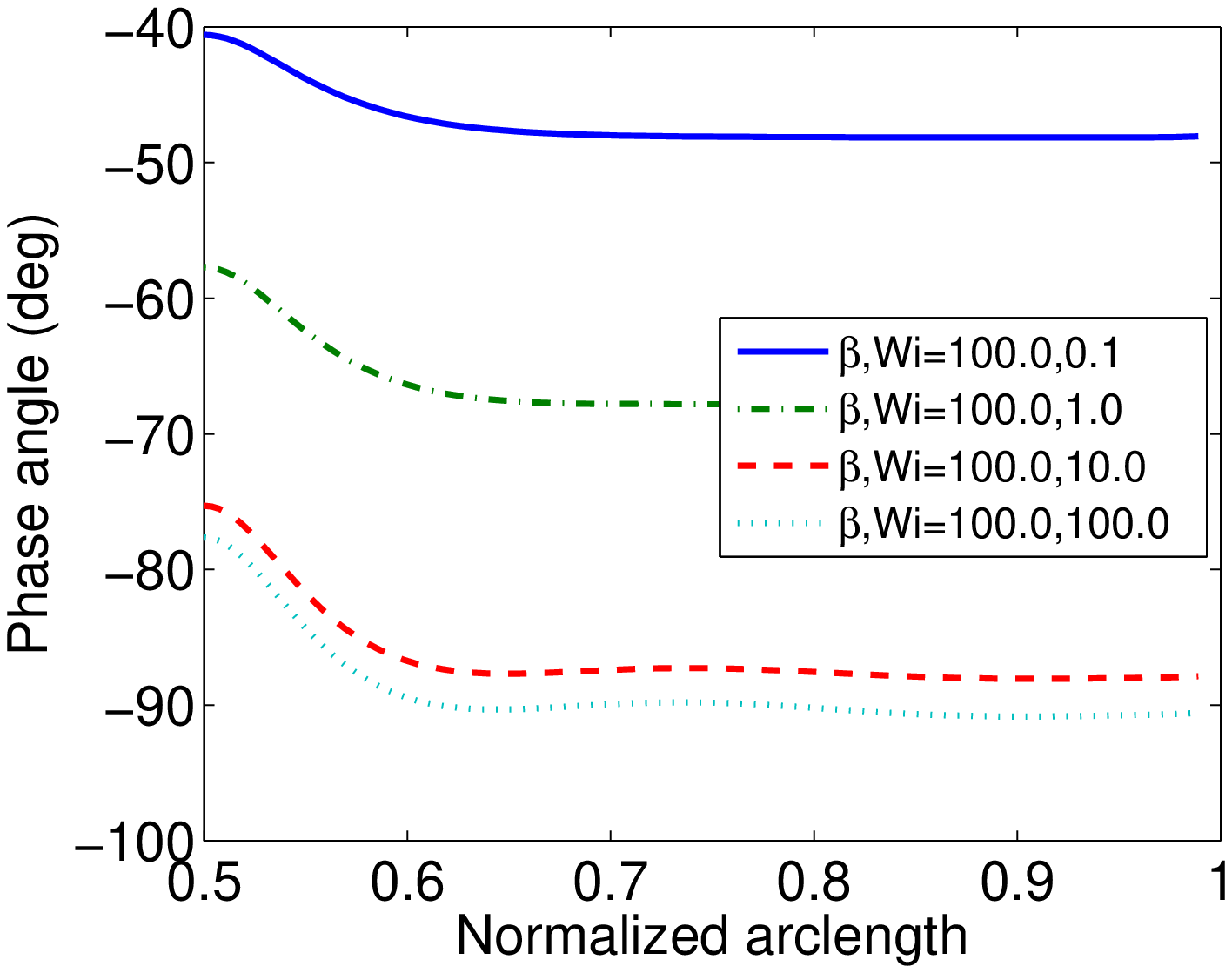}}
\caption[]{Amplitudes, normalized by its value at the midplane ($x=0$), and phases of shear stress on the oblate spheroid, plotted against the normalized arclength $s$, as the Weissenberg number $\Wi$ varies.}
\label{fig:obsnc}
\end{figure}

\begin{figure}
\centering
\subfigure[]{\includegraphics[height=1.8in]{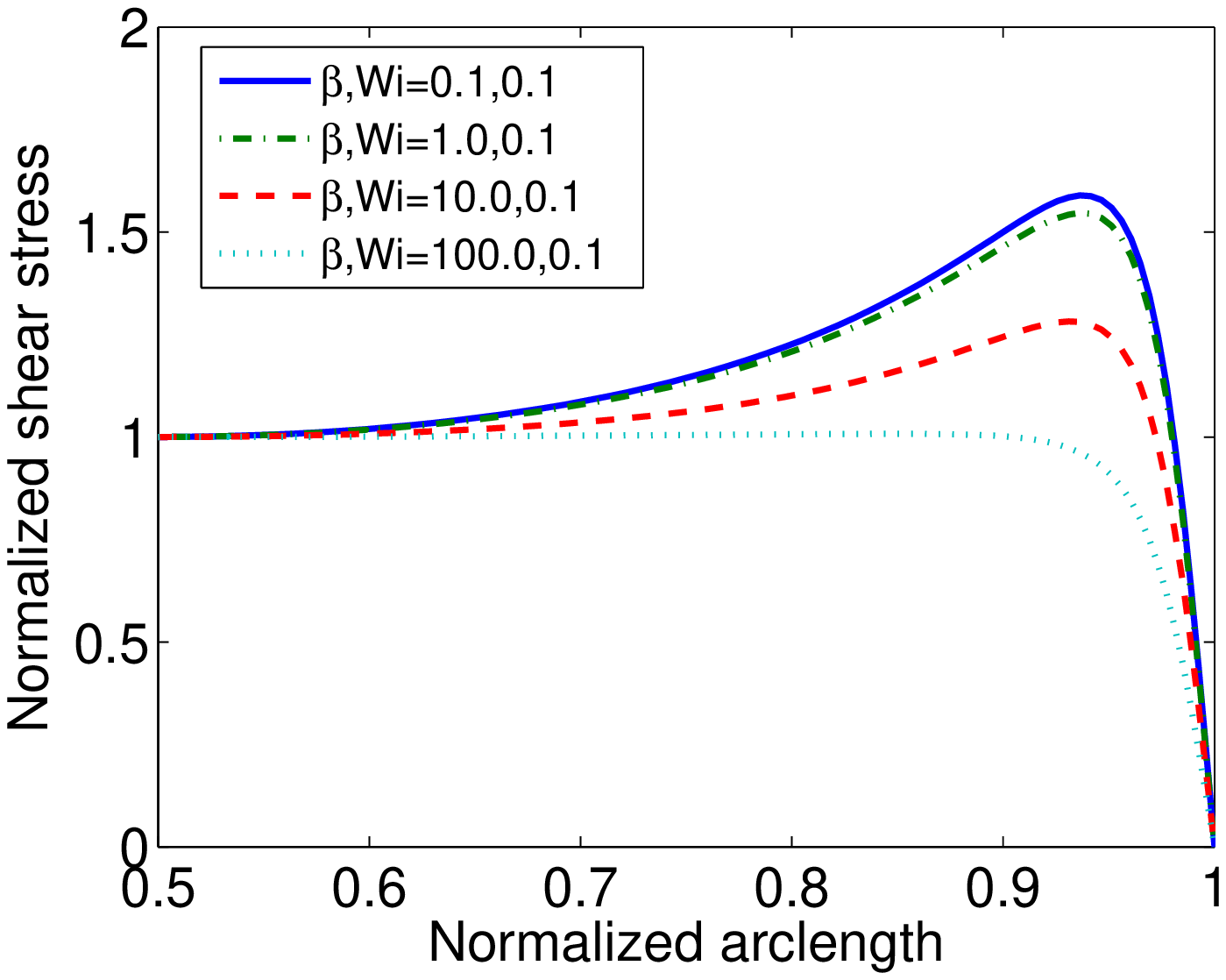}}
\subfigure[]{\includegraphics[height=1.8in]{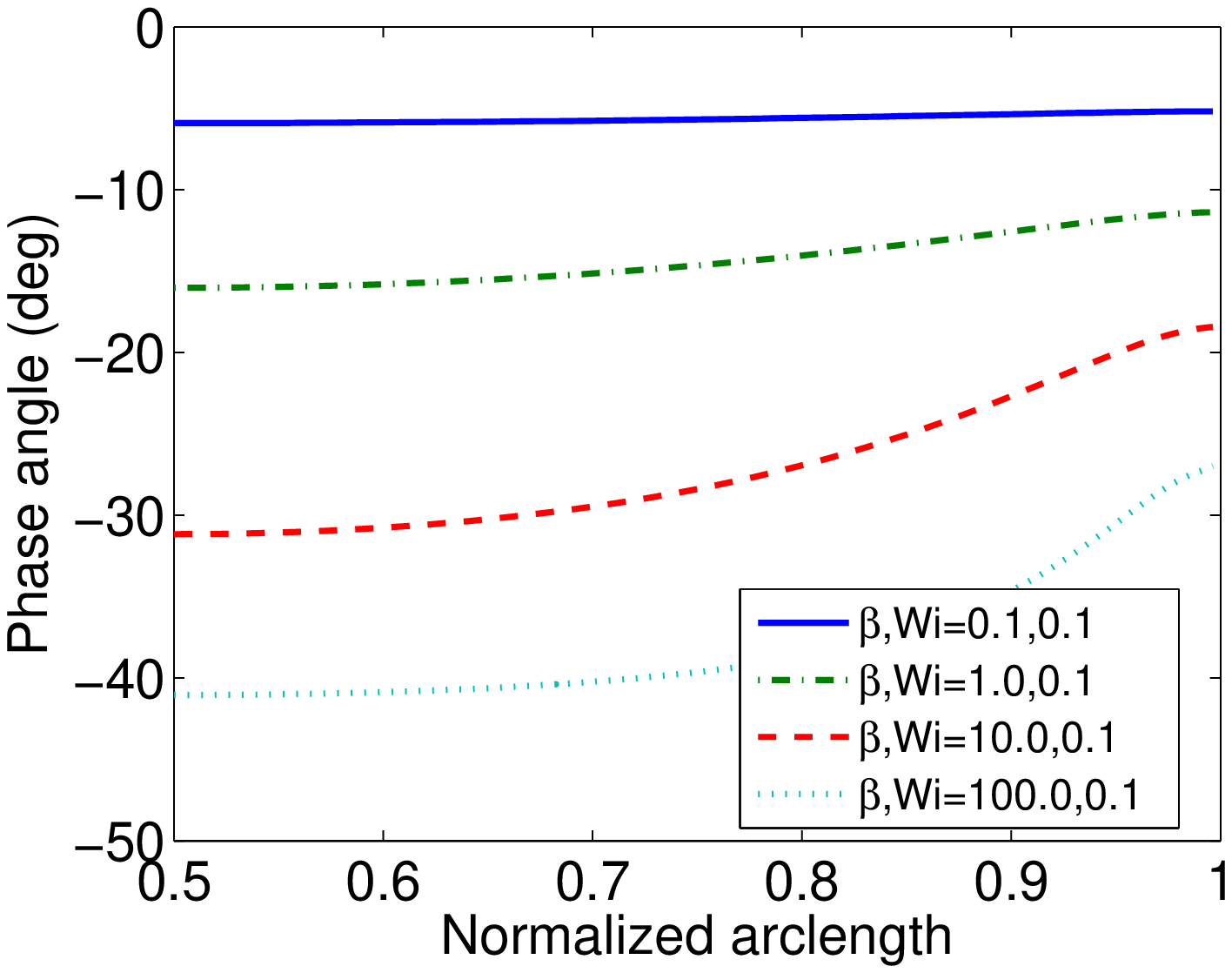}}
\caption[]{Amplitudes, normalized by its value at the mid-point ($x=0$), and phases of shear stress on the prolate spheroid, plotted against the normalized arclength $s$, as the frequency parameter $\beta$ varies. Weissenberg number $\Wi=0.1$.}
\label{fig:prosc0}
\end{figure}

Now we examine the shear stress on the spheroids, and we start with the case of oblate spheroids. The amplitudes and phases (not plotted if the stress is $0$) of the shear stress are plotted against the arclength at different frequency parameters and Weissenberg numbers, see Fig. \ref{fig:obscn} and Fig. \ref{fig:obsnc}. The arclength $s$ is normalized by total arclength of the particle contour in a meridional plane, with $s=0,0.5,1.0$ representing the right tip ($\sigma=0$), mid-point ($x=0$), and left tip ($\sigma=0$), respectively. Turning to the prolate spheroids, the amplitudes and phases of the shear stress behave similarly to those in the oblate spheroid case, with one difference. The stress achieves a maximum in magnitude near the tip ($\sigma=0$) instead of at the mid-point ($x=0$), when the frequency parameter is small enough. We plotted one case for the prolate spheroid to illustrate this difference with a varying frequency parameter and a Weissenberg number fixed at $\Wi=0.1$, see Fig. \ref{fig:prosc0}. We also checked the cases of higher Weissenberg numbers, and found that this behavior appears to be unaffected by change of the Weissenberg number. In contrast, the maximum stress always occurs at the mid-point in the oblate spheroid case.

As the frequency parameter $\beta$ increases beyond $O(10^2)$, the appearance of a boundary layer around the particle becomes evident. This oscillatory boundary layer, which appears at high frequency number, is to be differentiated from Prandtl's boundary layer, which is caused by viscous effects at high Reynolds number. It is a standard result in boundary layer theory \citep[sec. 5.13]{Batchelor_FD} that for oscillatory flow (Newtonian) along an infinite flat plate, the stress leads the velocity by a phase difference of $\theta=\pi/4$. This phenomenon can also be observed in our numerical results -- in low curvature, more flat, regions on the spheroids -- see Fig. \ref{fig:obscn}(b), where the flow is effectively Newtonian since $\Wi=0.1$ is sufficiently small. For oscillatory flow along an infinite flat plate in LVE, we can perform a standard small amplitude oscillatory shear (SAOS) analysis \cite{Faith_2001} to deduce that the stress leads the velocity by a phase difference of $\theta=\pi/4+\arctan(\Wi)/2$. This phenomenon can be observed in Fig. \ref{fig:obscn}(d), where the phase difference is $\theta\approx3\pi/8$ since $\Wi=1.0$, or in Fig. \ref{fig:obscn}(h), where the phase difference is $\theta\approx\pi/2$ since $\Wi=100.0$.

\subsection{Dumbbells and biconcave disks}
\begin{figure}
\centering
\includegraphics[height=1.6in]{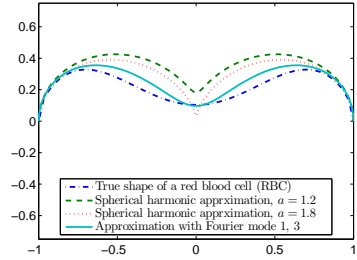}
\caption[]{Plots of the true shape, Eq. \eqref{rbcr}, of the red blood cell and its approximations Eq. \eqref{rbc1} with $a=1.2,1.8$, and Eq. \eqref{rbc2}, all normalized to the size of $1$.}
\label{fig:rbc_shape}
\end{figure}

Study of the flow around a biconcave disk under oscillatory motion is important for applications since for example, the red blood cell (RBC) is of this shape. We will also study the flow around a dumbbell as this shape also finds applications in engineering, e.g., particle agglomeration in suspensions. The biconcave disk or dumbbell shape can be modeled with a formula based on the spherical harmonic
\begin{eqnarray}
Y_2^0(\alpha,\phi)=\sqrt{\dfrac{5}{16\pi}}(3\cos^2\alpha-1),
\end{eqnarray}
where the coefficient is a normalization prefactor. For example, one can use
\begin{eqnarray}
r=1+a(3\cos^2\alpha-1)
\label{rbc1}
\end{eqnarray}
where $r=\sqrt{x^2+\sigma^2}$ is the polar distance and $\alpha$ is the polar angle. By varying the value of $a$ in $[-1,2]$, one can obtain a biconcave disk or dumbbell shape. However, we will use the following parametrization of the biconcave disk for better smoothness,
\begin{alignat}{3}
x&=(27\cos\alpha+4\cos3\alpha),\nonumber\\
\sigma&=(9\sin\alpha+6\sin3\alpha).
\label{rbc2}
\end{alignat}
This is obtained by using Fourier mode 1 and 3 to fit the average measurement (in Cartesian coordinates) of RBCs \cite{Evans_1980},
\begin{eqnarray}
y=d\sqrt{1-4\left(\dfrac{x}{d}\right)^2}\left[a_0+a_1\left(\dfrac{x}{d}\right)^2+a_2\left(\dfrac{x}{d}\right)^4\right],
\label{rbcr}
\end{eqnarray}
where $a_0=0.0518,a_1=2.0026,a_2=-4.491$, and $d=7.82\mu$m is the average diameter of an RBC. If better approximation is desired, mode 5 can be added. The formula Eq. \eqref{rbc1}, based on spherical harmonics, can be thought of as another parametrization of biconcave disks using the same Fourier modes ($1$ and $3$) with $\alpha$ being the polar angle, while $\alpha$ in Eq. \eqref{rbc2} is not exactly the polar angle. We have plotted the true shape of an RBC and its approximations in Fig. \ref{fig:rbc_shape}. It can be seen that Eq. \eqref{rbc2} provides a more accurate approximation to the true shape of an RBC and is more smooth.

\begin{figure}
\centering
\hskip -0.3in
\subfigure[]{\includegraphics[height=1.4in]{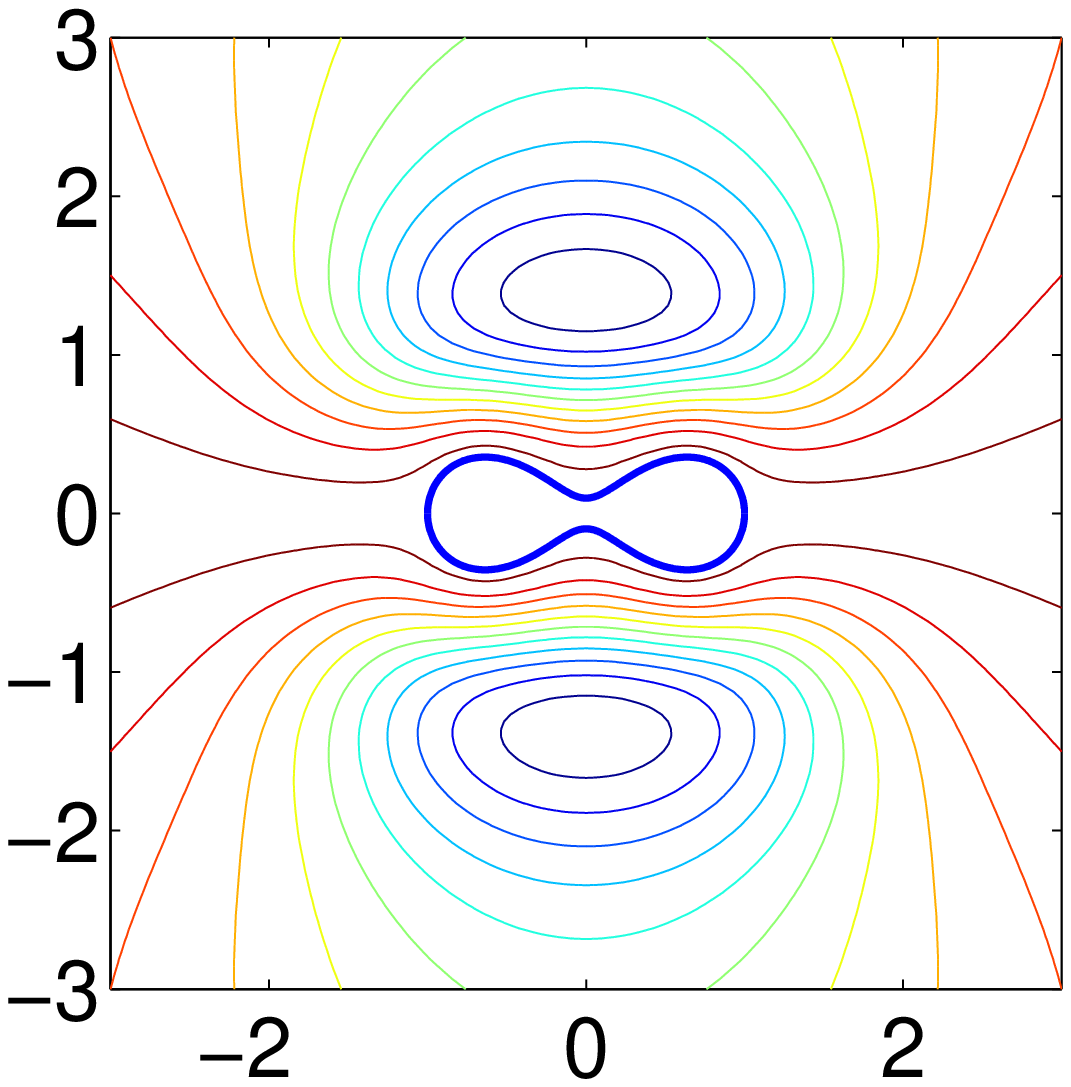}}
\hskip -0.3in
\subfigure[]{\includegraphics[height=1.4in]{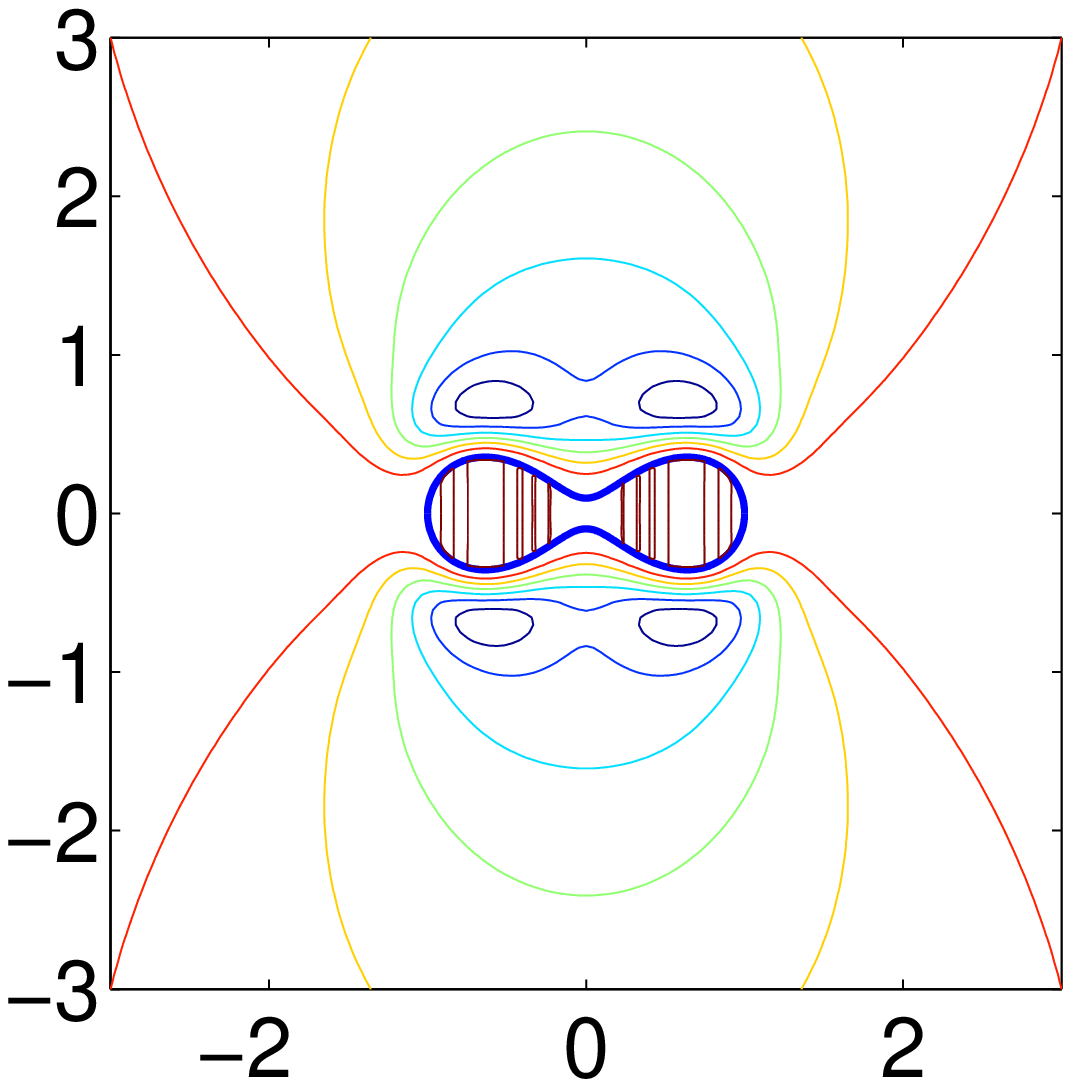}}
\hskip -0.3in
\subfigure[]{\includegraphics[height=1.4in]{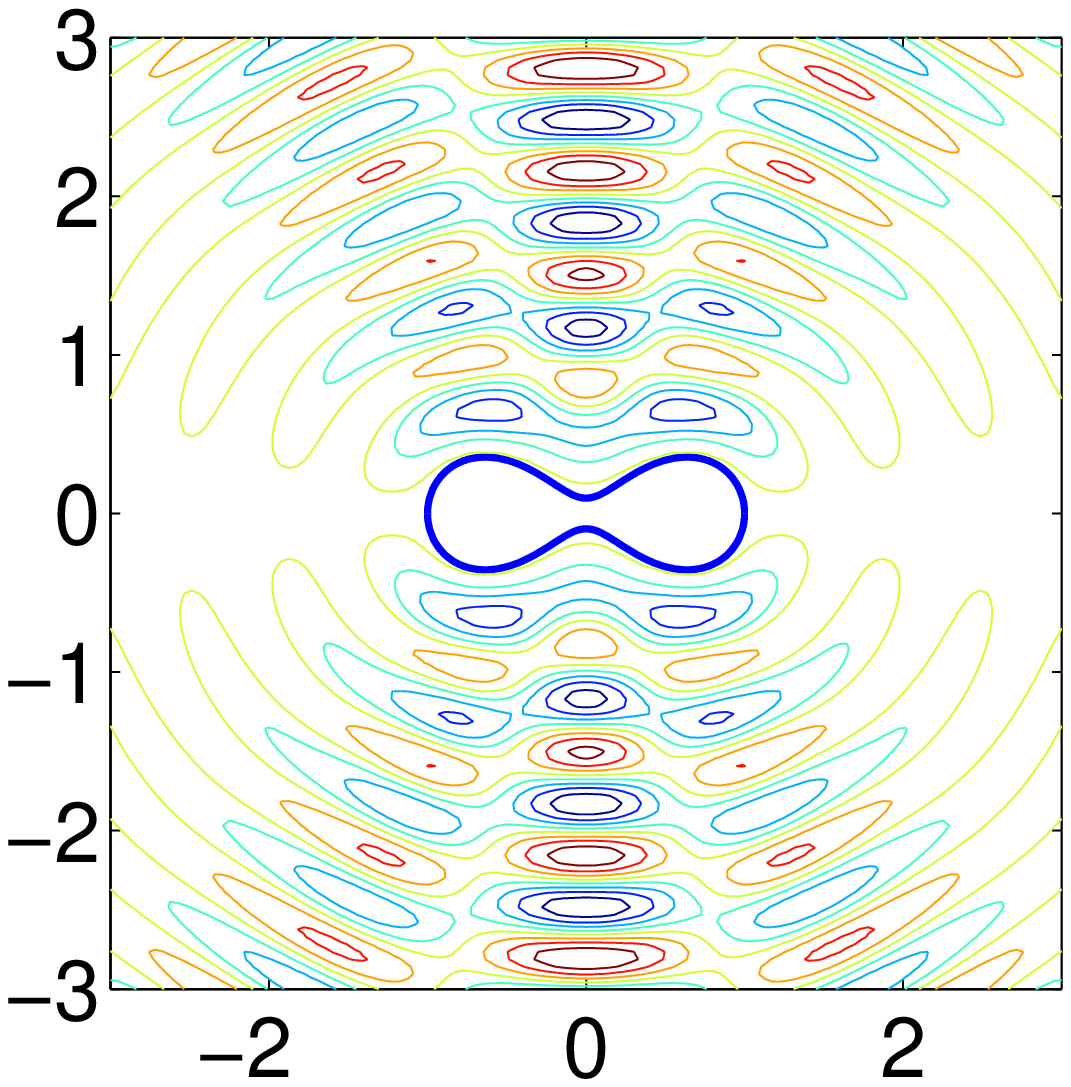}}
\caption[]{Streamlines around dumbbells at time $\omega t=0.5\pi$. (a) $\beta=10.0$ and $\Wi=0.1$. (b) $\beta=100.0$ and $\Wi=0.1$. (c) $\beta=100.0$ and $\Wi=100.0$.}
\label{fig:dumb_stream}
\end{figure}

We first look into the flow around the dumbbell, as shown in Fig. \ref{fig:dumb_stream}, when viscoelastic effects are not pronounced ($\Wi=0.1$). The streamlines are plotted at $\omega t=0.5\pi$, which is when the particle's velocity is 0 and starts accelerating. One can see that at a low frequency, Fig. \ref{fig:dumb_stream}(a), the flow conforms to the contour of the dumbbell and only one eddy is present. As the frequency increases, Fig. \ref{fig:dumb_stream}(b), the flow starts to feel the presence of a concave region and two eddies arise.
At high Weissenberg numbers, we can observe sequences of vortices, just like what we observed for the simpler geometry of a sphere. We plotted the streamlines around the dumbbell in Fig. \ref{fig:dumb_stream}(c) at $\beta=100.0$ and $\Wi=100.0$, which shows three streets of eddies.

\begin{figure}
\centering
\hskip -0.0in
\subfigure[]{\includegraphics[height=1.5in]{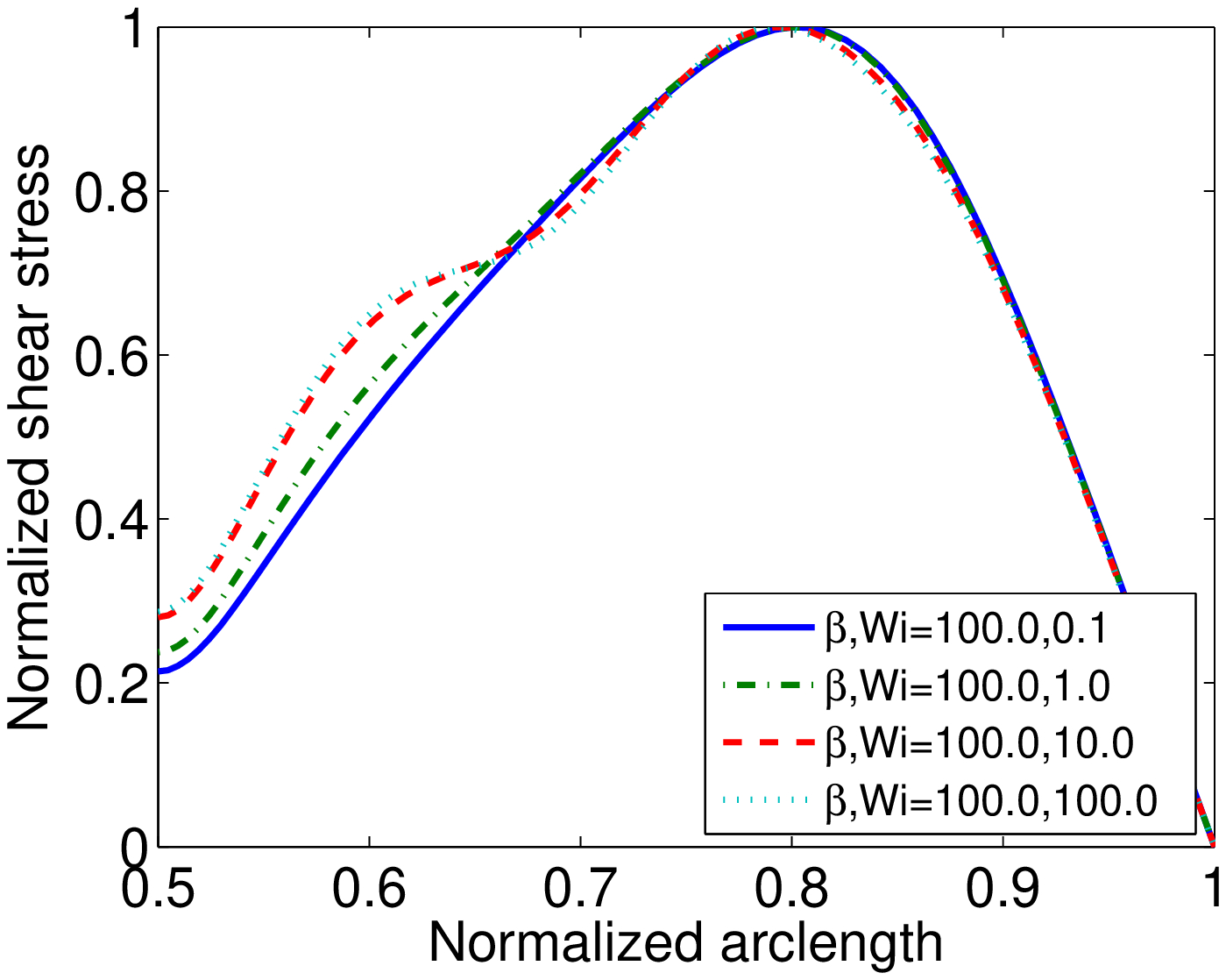}}
\hskip -0.0in
\subfigure[]{\includegraphics[height=1.5in]{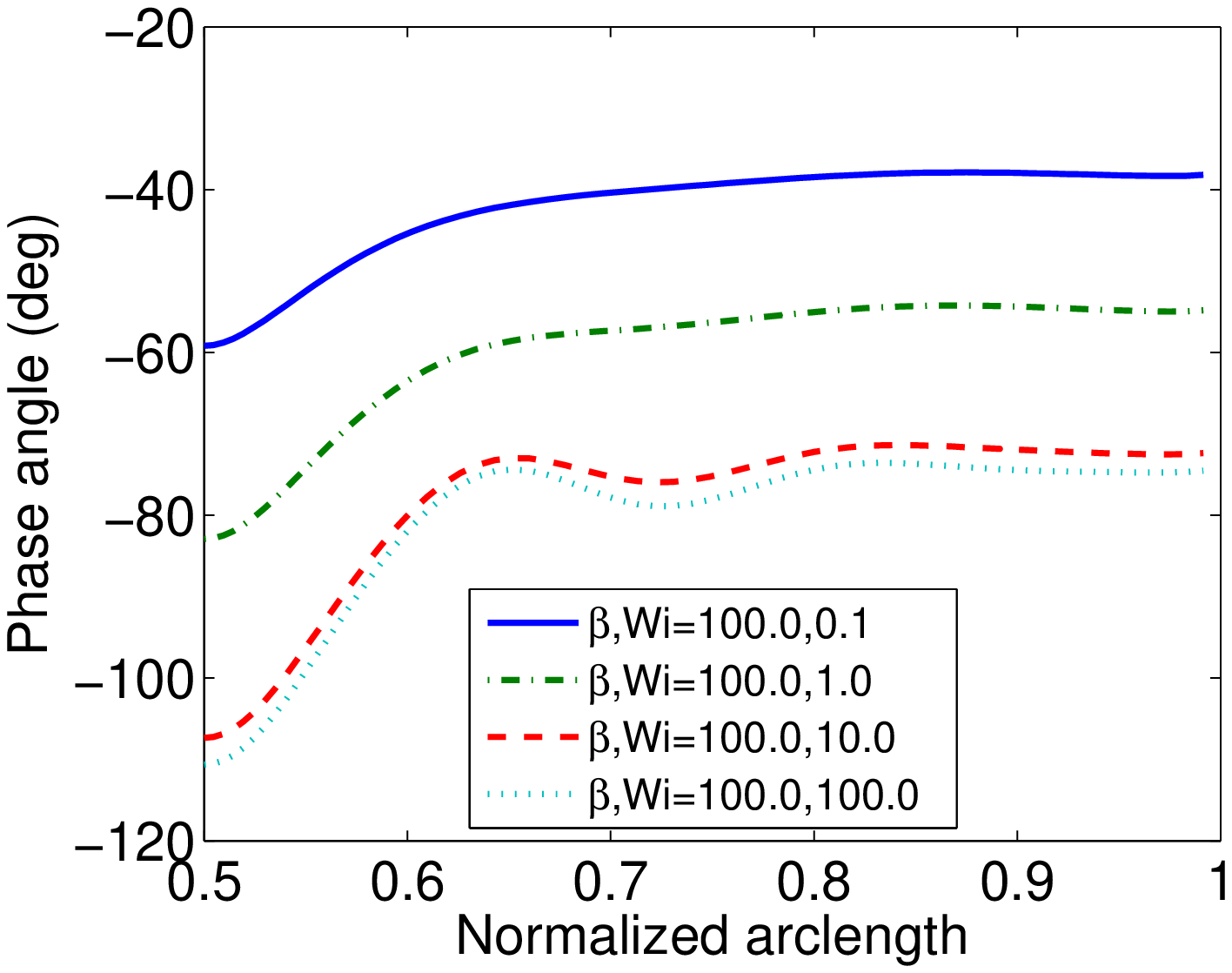}}
\caption[]{Amplitudes, normalized by its maximum value, and phases of shear stress on the dumbbell, plotted against the normalized arclength, at $\beta=100.0$ and as the Weissenberg number varies.}
\label{fig:dumb_strang}
\end{figure}

The modulus (normalized by its maximum value) and phase angle of the shear stress on the dumbbell at different Weissenberg numbers (with constant frequency number $\beta=100$) are plotted in Fig. \ref{fig:dumb_strang}.


\section{Conclusions}\label{sec.c}
We used a boundary integral method to study the disturbance flow caused by particles undergoing oscillatory motion in LVE, and demonstrated that the dynamics are generically different than those for the Newtonian case, in three aspects. First, there are streets of eddies, instead of only one as in the Newtonian case. Second, the eddies arise in the interior of the fluid and barely travel, while they are generated on particle surfaces and propagate into the fluid in the Newtonian case. Third, the appearance of eddies is delayed in comparison to the case of Newtonian flow, due to viscoelastic effects. These new discoveries should bear consequences on relevant engineering applications. We verified our numerical method by comparing its results to an existing analytic solution, in the simple case of fluid motion around one spherical particle. Mimicking a boundary layer theory for oscillatory flow along an infinite plate in Newtonian flow, we derived a similar (oscillatory) boundary layer theory for the case of LVE. We validated our numerical results against this new theory. We focused on two geometries because of their prevalence in applications: spheroids, and biconcave disks.\\

\bibliographystyle{jfm}

\bibliography{proposal,jds,indeiAP,Schieber_publications,visco,bibprop,Shuwangpub,XiaofanPub}

\end{document}